\documentclass[a4paper,11pt]{article}
\pdfoutput=1 % if your are submitting a pdflatex (i.e. if you have
\usepackage{jheppub} % for details on the use of the package, please
\usepackage[T1]{fontenc} % if needed
\usepackage[textsize=footnotesize,textwidth=2.5cm]{todo notes}

\definecolor{mikadoyellow}{rgb} {0.16, 0.254, 0.6}

\usepackage{amssymb,amsmath,latexsym,bm,amsfonts}
\usepackage{graphicx}
\usepackage{longtable}
\usepackage{color,xcolor}
\usepackage{indentfirst}
\usepackage{subfigure}
 \usepackage{comment}

\title{ \boldmath Entanglement contour and modular flow from subset entanglement entropies }

\author[a,b,c,d]{Qiang Wen}

\affiliation[a]{Shing-Tung Yau Center of Southeast University, Nanjing 210096, China}
\affiliation[b]{School of  Mathematics, Southeast University, Nanjing 211189, China}
\affiliation[c]{Graduate School, China Academy of Engineering Physics, Beijing 100193, China}
\affiliation[d]{Institute of Theoretical Physics, Chinese Academy of Sciences, Beijing 100190, China}

\emailAdd{wenqiang@seu.edu.cn}

\abstract{The Entanglement contour function quantifies the contribution from each degree of freedom in a region $\mathcal{A}$ to the entanglement entropy $S_{\mathcal{A}}$. Recently in \cite{Wen:2018whg} the author gave two proposals for the entanglement contour in two-dimensional theories. The first proposal is a fine structure analysis of the entanglement wedge, which applies to holographic theories. The second proposal is a claim that for general two-dimensional theories the partial entanglement entropy is given by a linear combination of entanglement entropies of relevant subsets inside $\mathcal{A}$. In this paper, we further study the partial entanglement entropy proposal by showing that it satisfies all the rational requirements proposed previously. We also extend the fine structure analysis from vacuum AdS space to BTZ black holes. Furthermore, we give a simple prescription to generate the local modular flows for two-dimensional theories from only the entanglement entropies without refer to the explicit Rindler transformations.}

\begin{document} 
\maketitle
\flushbottom

\section{Introduction}
The entanglement entropy for quantum many-body systems is defined as the von Neumann entropy of the reduced density matrix. It has become a quite hot topic in the study of modern theoretical physics.  In condense matter theories it can be used to distinguish new topological phases and characterize critical points, e.g., \cite{PhysRevLett.96.110405,Kitaev:2005dm,Calabrese:2004eu,Calabrese:2005zw,Hsu:2008af}. In high energy theory, in the context of AdS/CFT \cite{Maldacena:1997re,Witten:1998qj,Gubser:1998bc} the Ryu-Takayanagi (RT) formula \cite{Ryu:2006bv,Ryu:2006ef} (see also \cite{Song:2016pwx,Song:2016gtd,Jiang:2017ecm,Wen:2018mev} for its extension to holographies beyond AdS/CFT) relates quantum entanglement to spacetime geometry, thus entanglement entropy becomes an important tool to study quantum gravity and holography itself. 

However, entanglement entropy only contains part of the information in the reduced density matrix. The authors of \cite{Vidal} considered the possibility of the existence of a function $f_{\mathcal{A}}(i)$ which captures how much the degrees of freedom in site $i$ contribute to the entanglement entropy $S_{\mathcal{A}}$ of the region $\mathcal{A}$. In other words $f_{\mathcal{A}}(i)$, which we call the \textit{entanglement contour} function following \cite{Vidal}, characterizes the density function of the entanglement entropy in $\mathcal{A}$. In the continuum limit, we denote this function as $f_{\mathcal{A}}(x_1,\cdots x_{d-1})$, where $x_i$ with $1\leq i \leq d-1$, are the coordinates that parameterize $\mathcal{A}$, and $d$ is the spacetime dimension of the field theory. By definition it should satisfies the following basic requirements
\begin{align}\label{reqment}
S_{\mathcal{A}}=\int_{ \mathcal{A}}f_{\mathcal{A}}(x_1,\cdots x_{d-1})dx_1\cdots dx_{d-1}\,,\qquad
f_{\mathcal{A}}(x_1,\cdots x_{d-1})\geq 0\,.
\end{align}
Also in \cite{Vidal}, a set of physical requirements for the entanglement contour is proposed. However these requirements cannot uniquely determine the contour function. The fundamental definition of the entanglement contour based on the reduced density matrix is still not clear. We will indirectly study the contour function $f_{\mathcal{A}}(x)$ through the \textit{partial entanglement entropy} (PEE) $s_{\mathcal{A}}(\alpha)$ for any subset $\alpha$ of $\mathcal{A}$, which captures the contribution from $\alpha$ to $S_{\mathcal{A}}$. The PEE is defined by
\begin{align}\label{a2definition}
s_{\mathcal{A}}(\alpha)=\int_{ \alpha}f_{\mathcal{A}}(x_1,\cdots x_{d-1})dx_1\cdots dx_{d-1}\,.
\end{align}
Without loss of generality, we only consider $\alpha$ to be connected. The entanglement contour can be reproduced from the PEE. 

Though our understanding of the concept of PEE is primitive, there are already several proposals to construct the contour functions. In Ref.\cite{Wen:2018whg} it was proposed that the PEE of a subset can be given by a linear combination of the entanglement entropies of certain subsets in $\mathcal{A}$. We call it the PEE \textit{proposal}. Also in Ref.\cite{Wen:2018whg} the author find a natural slicing of the entanglement wedges in Poincar\'e AdS$_3$ by the so-called modular planes. This \textit{fine structure analysis} of the entanglement wedge gives a fine correspondence between the points on the boundary interval and the points on the corresponding RT surface, from which we can read the entanglement contour. In this paper we will show that the PEE \textit{proposal} satisfies all the physical requirements proposed in \cite{Vidal}. We also extend the \textit{fine structure analysis} of the entanglement wedge from the vacuum AdS$_3$ to the BTZ backgrounds. In this case the \textit{fine structure analysis} gives the same entanglement contour function as the PEE \textit{proposal}.  The \textit{fine structure analysis} indicates that the PEE is invariant under the modular flow. Based on this property we propose a simple prescription to generate the local modular flows from only the entanglement entropies in two-dimensional theories. Based on the Rindler method \cite{Casini:2011kv,Song:2016gtd,Jiang:2017ecm}, the modular flows for several holographic theories were previously carried out in \cite{Jiang:2017ecm,Wen:2018mev,Jiang:2019qvd}. Our prescription can reproduce the modular flows in a much easier way without refer to the Rindler transformations.

Previously the attempts to construct the entanglement contour functions are based on the reduced density matrix and confined to the Gaussian states in free theories \cite{Botero,Vidal,PhysRevB.92.115129,Coser:2017dtb,Tonni:2017jom}. We may call this construction the \textit{Gaussian formula} (see \cite{DiGiulio:2019lpb,Kudler-Flam:2019nhr} for more recent developments along this line). The consistency check between the \textit{Gaussian formula} and the PEE \textit{proposal} for some cases of free boson and free fermion can be found in \cite{Kudler-Flam:2019nhr}.  The entanglement contour gives a finer description for the entanglement structure and has been shown to be particularly useful when studying the dynamical situations \cite{Vidal,Kudler-Flam:2019oru,DiGiulio:2019lpb}. Also it has been recently demonstrated that the entanglement contour (calculated by the PEE \textit{proposal}) is an useful probe of slowly scrambling and non-thermalizing dynamics for some interacting many-body systems \cite{MacCormack:2020auw}. The holographic picture of entanglement contour \cite{Wen:2018whg,Wen:2018mev} should be closely related to the other holographic formalisms that can give a finer description of holographic entanglement, like the tensor network \cite{Swingle:2009bg} and the bit threads picture \cite{Freedman:2016zud} (for related discussions see \cite{Kudler-Flam:2019oru,Han:2019scu}). We expect the new concept of entanglement contour in quantum information to play an important role in gauge/gravity dualities, entanglement structure of quantum field theories and condense matter theories.

\section{The partial entanglement entropy proposal}

\begin{figure}[h] 
   \centering
    \includegraphics[width=0.4 \textwidth]{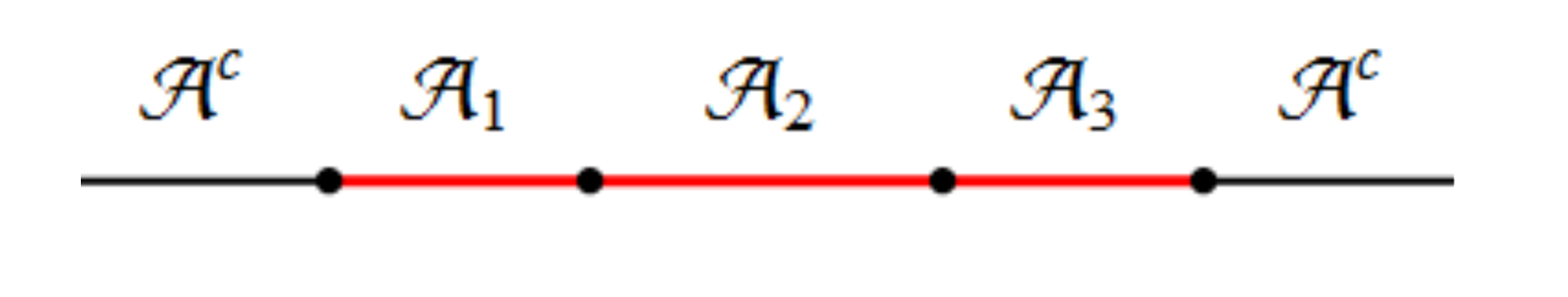} 
 \caption{The partition of an single interval, $\mathcal{A}=\mathcal{A}_1\cup\mathcal{A}_2\cup\mathcal{A}_3$. 
\label{partition} }
\end{figure}

In this paper, we focus on the entanglement contour functions for single intervals in 2-dimensional theories. Let us consider the PEE of an arbitrary subset $\alpha$ of $\mathcal{A}$, which in general divide $\mathcal{A}$ into three subsets, $\mathcal{A}=\alpha_L\cup \alpha\cup \alpha_R$. Here $\alpha_L$ and $\alpha_R$ represent the subsets on the left and right hand side of $\alpha$. The subsets $\alpha_L$ or $\alpha_R$ vanishes if $\alpha$ shares one boundary with $\mathcal{A}$, hence the partition only involves two subsets. Also we can choose $\alpha=\mathcal{A}$ hence both of $\alpha_L$  and $\alpha_R$ vanish. The PEE \textit{proposal} gives an \textit{ansatz} for the PEE $s_{\mathcal{A}}(\alpha)$, which is a linear combination of subset entanglement entropies. Let, for example, $S_{\alpha\cup \alpha_L}$ denote the entanglement entropy of the subset $\alpha\cup \alpha_L$, the proposal claims,
\begin{align}\label{ECproposal}
s_{\mathcal{A}}(\alpha)=\frac{1}{2}\left(S_{\alpha\cup \alpha_L}+S_{\alpha\cup \alpha_R}-S_{ \alpha_L}-S_{\alpha_R}\right)\,.
\end{align}
Consider the case shown in Fig.\ref{partition}, if we chose $\alpha=\mathcal{A}_2$, then we have $\alpha_L=\mathcal{A}_1$ and $\alpha_R=\mathcal{A}_3$. The \textit{proposal} \eqref{ECproposal} gives the $s_{\mathcal{A}}(\mathcal{A}_2)$ in \cite{Wen:2018whg} 
\begin{align}\label{ECproposal2}
s_{\mathcal{A}}(\mathcal{A}_2)=\frac{1}{2}\left(S_{\mathcal{A}_1\cup\mathcal{A}_2}+S_{\mathcal{A}_2\cup\mathcal{A}_3}-S_{\mathcal{A}_1}-S_{\mathcal{A}_3}\right)\,.
\end{align}
We can also apply the proposal \eqref{ECproposal} to $\mathcal{A}_1$ and $\mathcal{A}_3$. For example, if we consider $\alpha=\mathcal{A}_1$, then we should have $\alpha_R=\mathcal{A}_2\cup\mathcal{A}_3$ while $\alpha_L$ vanishes. So according to \eqref{ECproposal} we should have
\begin{align}\label{ECproposal1}
s_{\mathcal{A}}(\mathcal{A}_1)=\frac{1}{2}\left(S_{\mathcal{A}_1}+S_{\mathcal{A}}-S_{\mathcal{A}_2\cup\mathcal{A}_3}\right)\,.
\end{align}
Similarly we can easily get
\begin{align}\label{ECproposal3}
s_{\mathcal{A}}(\mathcal{A}_3)=\frac{1}{2}\left(S_{\mathcal{A}_3}+S_{\mathcal{A}}-S_{\mathcal{A}_1\cup \mathcal{A}_2}\right)\,.
\end{align}
For consistency, one can easily check that
\begin{align}
s_{\mathcal{A}}(\mathcal{A}_1)+s_{\mathcal{A}}(\mathcal{A}_2)+s_{\mathcal{A}}(\mathcal{A}_3)=S_{\mathcal{A}}\,.
\end{align}

Since the PEE \textit{proposal} \eqref{ECproposal} is a linear combination of subset entanglement entropies, it will be interesting to explore the properties satisfied by the \textit{proposal} using the inequalities \cite{araki1970,Headrick:2013zda} obeyed by entanglement entropy. In the following we list some of the important properties satisfied by $s_{\mathcal{A}}(\alpha)$:

\begin{enumerate}
\item \textbf{Additivity}: Splitting an arbitrary subset $\mathcal{A}_i$ into arbitrary two adjacent parts $\mathcal{A}_{i}=\mathcal{A}_{i}^{a}\cup \mathcal{A}_{i}^{b}$, we can choose $\alpha=\mathcal{A}_{i},\,$ $\mathcal{A}_{i}^{a}$ and $\mathcal{A}_{i}^{b}$ respectively to find that
\begin{align}\label{additivity}
s_{\mathcal{A}}(\mathcal{A}_{i})&=s_{\mathcal{A}}(\mathcal{A}_{i}^{a})+s_{\mathcal{A}}(\mathcal{A}_{i}^{b})\,,
\end{align}

\item \textbf{Normalization}: It is straightforward to see that when $\alpha=\mathcal{A}$ we should have
\begin{align}
 S_{\mathcal{A}}&=s_{\mathcal{A}}(\alpha)|_{\alpha=\mathcal{A}}\,.
\end{align}

\item \textbf{Positivity}: The strong subadditivity \cite{Lieb:1973cp,Lieb:1973zz} directly gives $s_{\mathcal{A}}(\alpha)>0$.

\item \textbf{Invariance under local transformations}: Since the entanglement entropy of a region is invariant under local transformations that only act on this region, by definition $s_{\mathcal{A}}(\alpha)$ is invariant under local transformations which act only inside $\alpha_L$, $\alpha$ and $\alpha_R$ respectively.

\item \textbf{Upper bound}: Since in general we have $S_{\alpha_L}+S_{\alpha}\geq S_{\alpha\cup \alpha_L}$ and $S_{\alpha}+S_{\alpha_R}\geq S_{\alpha\cup \alpha_R}$, together with \eqref{ECproposal} we have 
\begin{align}\label{upperbound}
 s_{\mathcal{A}}(\alpha) \leq S_{\alpha} \,.
\end{align}
The upper bound \eqref{upperbound} indicates that for any site or subset $\alpha$ inside $\mathcal{A}$,  $s_{\mathcal{A}}(\alpha)$ should not be larger than the entanglement entropy of the subset. This is reasonable when we interpret $s_{\mathcal{A}}(\alpha)$ as the contribution from $\alpha$ to the entanglement entropy $S_{\mathcal{A}}$. Because, unlike $S_{\alpha}$, $s_{\mathcal{A}}(\alpha)$ does not count the entanglement between $\alpha$ and other subsets inside $\mathcal{A}$, thus should be no larger.

\item \textbf{Symmetry}: Given region $\mathcal{A}$ and a partition into the subsets $\{\mathcal{A}_i\}$, we consider a symmetry transformation $\mathcal{T}$ of the theory thus $\mathcal{T} \mathcal{A}= \mathcal{A}'$ and $\mathcal{T} \mathcal{A}_{i}= \mathcal{A}'_{i}$. Since $\mathcal{T}$ is a symmetry, the subsets $\mathcal{A}_i$ and $\mathcal{A}'_i$ should play the equivalent role in $\mathcal{A}$. As a result we should have $S_{\mathcal{A}_i}=S_{\mathcal{A}'_i}$, and furthermore  $S_{\mathcal{A}_i\cup\mathcal{A}_j}=S_{\mathcal{A}'_i\cup\mathcal{A}'_j}$. This means
\begin{align}
\frac{1}{2}\left(S_{\alpha\cup \alpha_L}+S_{\alpha\cup \alpha_R}-S_{ \alpha_L}-S_{\alpha_R}\right)
=\frac{1}{2}\left(S_{\alpha'\cup \alpha'_L}+S_{\alpha'\cup \alpha'_R}-S_{ \alpha'_L}-S_{\alpha'_R}\right)\,,
\end{align}
thus
\begin{align}\label{symmetry}
s_{\mathcal{A}}(\alpha)=s_{\mathcal{A}'}(\alpha')\,.
\end{align}
In other words the PEE \textit{proposal} respect the symmetry $\mathcal{T}$.
\end{enumerate}

For holographic CFTs, the \textit{proposal} furthermore satisfies a lower bound because the mutual information is monogamous \cite{Hayden:2011ag}, i.e. 
\begin{align}\label{monogamy}
S_{\mathcal{A}_{1}}+S_{\mathcal{A}_{2}}+S_{\mathcal{A}_{3}}+S_{\mathcal{A}}\leq
 S_{\mathcal{A}_{1}\cup\mathcal{A}_{2}}+S_{\mathcal{A}_{2}\cup\mathcal{A}_{3}}+S_{\mathcal{A}_{1}\cup\mathcal{A}_{3}}\,.
\end{align}
The above equality indicates that
\begin{align}\label{lowerbound}
s_{\mathcal{A}}(\mathcal{A}_2)\geq  \frac{1}{2}\left( S_{\mathcal{A}}-(S_{\mathcal{A}_{1}\cup\mathcal{A}_{3}}-S_{\mathcal{A}_{2}})\right) \geq   0\,,
\end{align}
where the second inequality comes from the Araki-Lieb triangle inequality $ S_{\mathcal{A}} \geq |S_{\mathcal{A}_{1}\cup\mathcal{A}_{3}}-S_{\mathcal{A}_{2}})|$.

We can test its rationality using the Bit thread picture \cite{Freedman:2016zud}, which quantifies the quantum entanglement with a set of ``Bit threads'' with a cross-sectional area of 4 Planck areas and can only end on the boundary or the horizon. In this picture $S_{\mathcal{A}}$ is the maximum possible number of threads emanating from $\mathcal{A}$ and end on $\mathcal{A}^{c}$. Also the PEE $s_{\mathcal{A}}(\mathcal{A}_2)$ have a quite natural definition as the number of threads emanating from $\mathcal{A}_2$ and end on $\mathcal{A}^{c}$, when the number of threads from $\mathcal{A}$ to $\mathcal{A}^{c}$ is maximal. So far all the configurations that maximize the flux of bit threads on the RT surface $\mathcal{E}_\mathcal{A}$ are considered to be degenerate in the bit thread picture as they give the same $S_{\mathcal{A}}$. However different degenerate configurations will give different $s_{\mathcal{A}}(\mathcal{A}_2)$ thus the entanglement contour can be different. In the bit thread picture $s_{\mathcal{A}}(\mathcal{A}_2)$ has upper and lower bounds. For example, $s_{\mathcal{A}}(\mathcal{A}_2)$ reaches its upper bound when the number of bit threads emanating from $\mathcal{A}_2$ is maximized and they all cross $\mathcal{E}_{\mathcal{A}}$. So we have $s_{\mathcal{A}}(\mathcal{A}_2)\leq S_{\mathcal{A}_2}$, which coincide with \eqref{upperbound}. Similarly we have $s_{\mathcal{A}}(\mathcal{A}_1\cup\mathcal{A}_3)\leq S_{\mathcal{A}_1\cup\mathcal{A}_3}$. On the other hand when $s_{\mathcal{A}}(\mathcal{A}_1\cup\mathcal{A}_3)$ reaches its upper bound $s_{\mathcal{A}}(\mathcal{A}_2)$ will reach its lower bound, which is given by
\begin{align}\label{lowerboundbt}
s_{\mathcal{A}}(\mathcal{A}_2)\geq S_{\mathcal{A}}-S_{\mathcal{A}_1\cup\mathcal{A}_3}\,.
\end{align}
The above inequality makes sense when $\mathcal{A}_1$ or $\mathcal{A}_3$ vanishes thus the right and side can be positive. It is interesting that, since $S_{\mathcal{A}_2}\geq S_{\mathcal{A}}-S_{\mathcal{A}_1\cup \mathcal{A}_3}$, the lower bound \eqref{lowerbound} for $s_{\mathcal{A}}(\mathcal{A}_2)$ defined by \eqref{ECproposal} is stronger than the bound \eqref{lowerboundbt} we get from the bit thread picture. This is not surprising since the bit threads have the largest freedom to move in the bulk thus should give the weakest bound for the PEE. This implies the freedom for the bit threads may be confined in some way.

It is also interesting to compare the PEE \textit{proposal} with mutual information. Assuming $\mathcal{A}^{c}$ is the region that purifies $\mathcal{A}$ hence $\mathcal{A}\cup\mathcal{A}^{c}$ is in a pure state. When the subset $\mathcal{A}_2$ shares one boundary with $\mathcal{A}$, it is easy to see that the $s_{\mathcal{A}}(\mathcal{A}_2)$ calculated by the \textit{proposal} is just half of the mutual entanglement entropy between $\mathcal{A}_2$ and $\mathcal{A}^{c}$,
\begin{align}\label{a13mutual}
s_{\mathcal{A}}(\mathcal{A}_2)=\frac{1}{2} I(\mathcal{A}_2:\mathcal{A}^{c})\,.
\end{align}
Similarly, in the case of Fig.\ref{partition}, we have $s_{\mathcal{A}}(\mathcal{A}_{1}\cup\mathcal{A}_2)=\frac{1}{2}  I(\mathcal{A}_1\cup\mathcal{A}_2:\mathcal{A}^{c})$, thus
\begin{align}
s_{\mathcal{A}}(\mathcal{A}_2)&=\frac{1}{2} \left( I(\mathcal{A}_1\cup\mathcal{A}_2:\mathcal{A}^{c})- I(\mathcal{A}_1:\mathcal{A}^{c})\right)\,.
\end{align}
For holographic CFT, the mutual information is monogamous, so $I(\mathcal{A}_1\cup\mathcal{A}_2:\mathcal{A}^{c})\geq I(\mathcal{A}_1:\mathcal{A}^{c})+I(\mathcal{A}_2:\mathcal{A}^{c})$. Using $S_{\mathcal{A}}=S_{\mathcal{A}^{c}}$ and $S_{\mathcal{A}_1\cup\mathcal{A}_3}=S_{\mathcal{A}_{2}\cup \mathcal{A}^{c}}$, we can write the lower bound \eqref{lowerbound} as
\begin{align}\label{pemi2}
s_{\mathcal{A}}(\mathcal{A}_2)\geq  \frac{1}{2}\left( S_{\mathcal{A}^{c}}+S_{\mathcal{A}_{2}}-S_{\mathcal{A}_{2}\cup\mathcal{A}^{c}}\right) =\frac{1}{2} I(\mathcal{A}_2:\mathcal{A}^{c})\,.
\end{align}
This means the $s_{\mathcal{A}}(\mathcal{A}_2)$ calculated by the \textit{proposal} is always larger than half of the mutual information between $\mathcal{A}_2$ and $\mathcal{A}^{c}$. 

Though the PEE \textit{proposal} \eqref{ECproposal} satisfies all the physical requirements proposed in \cite{Vidal}, it is not justified because those requirements are not enough to uniquely determine the PEE. More generally, assuming that the PEE is a linear combination of all the subset entanglement entropies\footnote{This is similar to the strategy used in \cite{Hubeny:2007re} to derive the entanglement entropies for two disconnected intervals. However in \cite{Hubeny:2007re} the linear combination that gives the entanglement entropy undergoes a phase transition thus the derivation is flawed.}, i.e.
\begin{align}\label{ansatz}
s_{\mathcal{A}}(\alpha) =&c_1 S_{\alpha_{L}}+c_2 S_{\alpha}+c_3 S_{\alpha_R}+c_{12}S_{\alpha_L\cup \alpha}
+c_{23}S_{\alpha\cup \alpha_R}+c_{13} S_{\alpha_L\cup \alpha_R}+c_{123}S_{\mathcal{A}}\,，
 \end{align} 
Where the coefficients are constants that are independent from the choice of the partition. Imposing the requirements of additivity \eqref{additivity} and normalization, one can uniquely determine the coefficients in the ansatz and get \eqref{ECproposal}.

\section{Fine structure analysis in BTZ black holes}
\subsection{The fine correspondence in BTZ background}
In the context of AdS$_3$/CFT$_2$, a holographic picture for the entanglement contour of a single interval is given in \cite{Wen:2018whg}. When the modular flow is local, the boundary modular flow lines in the causal development $\mathcal{D}_{\mathcal{A}}$ can be defined as the integral curves along the boundary modular flow. Allowing the points on a modular flow line to flow under the bulk modular flow, the trajectory will form a two-dimensional slice in the entanglement wedge. We call these slices the \textit{modular planes}. From this construction, we can consider the entanglement wedge as a slicing of the modular planes. Each modular plane will intersect with the interval $\mathcal{A}$ on a point $\mathcal{A}(\bar{x}_m)$ and its RT surface $\mathcal{E}_{\mathcal{A}}$ on another point $\mathcal{E}_{\mathcal{A}}(x_m)$. It was argued in \cite{Wen:2018whg} that the cyclic gluing of the point $\mathcal{A}(\bar{x}_m)$ turns on the nonzero contribution to the $S_{\mathcal{A}}$ on $\mathcal{E}_{\mathcal{A}}(x_m)$, thus gives a fine correspondence between the points on $\mathcal{A}$ and $\mathcal{E}_{\mathcal{A}}$.  We can read the entanglement contour function from this fine correspondence. The discussion in \cite{Wen:2018whg} is restricted to the pure AdS$_3$ space. We stress that the prescription can be applied to more general cases with local modular flow. Here we extend the discussion to BTZ black holes.

The key to conduct the \textit{fine structure analysis} of the entanglement wedge is the bulk and boundary modular flows. In the case of pure AdS$_3$, a Rindler transformation, which is a symmetry transformation, can be constructed to map the causal development $\mathcal{D}_{\mathcal{A}}$ to a ``Rindler space'' with infinitely far away boundaries. This Rindler transformation can be extend to the bulk \cite{Song:2016gtd} which map the corresponding entanglement wedge to a Rindler AdS$_3$ space, which is the exterior region of a hyperbolic black hole \cite{Casini:2011kv},
\begin{align}\label{hyperbolic}
ds^2=\frac{d\tilde{r}^2}{\tilde{r}^2/L^2-1}-(\tilde{r}^2/L^2-1)d\tilde{\tau}^2+\tilde{r}^2 d\tilde{x}^2\,.
\end{align}
The horizon $\tilde{r}=L$ of the hyperbolic black hole maps to the RT surface in the oringinal AdS$_3$, and the Rindler time translation $\partial_{\tilde{\tau}}$ maps to the bulk modular flow in the entanglement wedge. So with the inverse bulk Rinlder transformation we can solve the explicit information for bulk and boundary modular flows, and furthermore construct the modular planes in the entanglement wedge \cite{Wen:2018whg}.

In principle the above prescription can be extend to the BTZ black holes. In other words we can also try to construct a bulk Rindler transformation map the entanglement wedge in BTZ background to the hyperbolic black hole \eqref{hyperbolic}. However, the construction of the Rindler transformations will be much more complicated. The explicit bulk modular flow in this background is obtained in \cite{Jiang:2019qvd} using other tricks. Constructing modular planes directly in the entanglement wedge following the bulk and boundary modular flows seems to be tough and unwise. However, it will be much easier to do the \textit{fine structure analysis} in the Rindler bulk space. The images of the static interval $\mathcal{A}$ and its RT surface $\mathcal{E}_{\mathcal{A}}$ (at the zero time slice) in the Rindler bulk \eqref{hyperbolic} are given by
\begin{align}
\tilde{\mathcal{A}}:\{\tilde{\tau}=0,\quad \tilde{r}=\infty\}\,,\qquad \tilde{\mathcal{E}}_{\tilde{\mathcal{A}}}:\{\tilde{\tau}=0,\quad \tilde{r}=L\}\,.
\end{align}
Our goal is to find the fine correspondence between points on $\tilde{\mathcal{A}}$ and $\tilde{\mathcal{E}}_{\tilde{\mathcal{A}}}$ and then map this fine correspondence back to the original BTZ black hole.

\begin{figure}[h] 
   \centering
    \includegraphics[width=0.35 \textwidth]{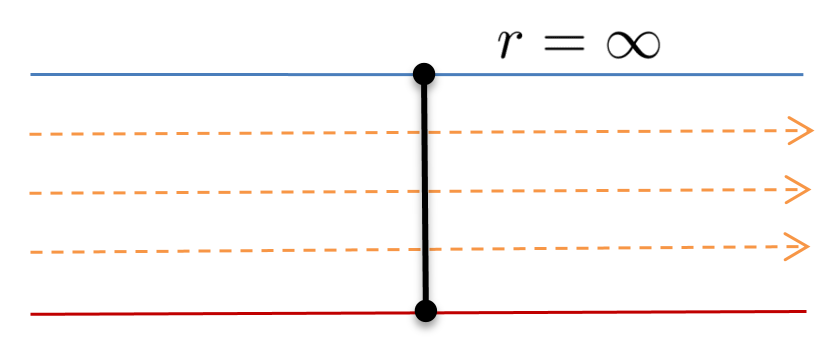}\qquad
        \includegraphics[width=0.35 \textwidth]{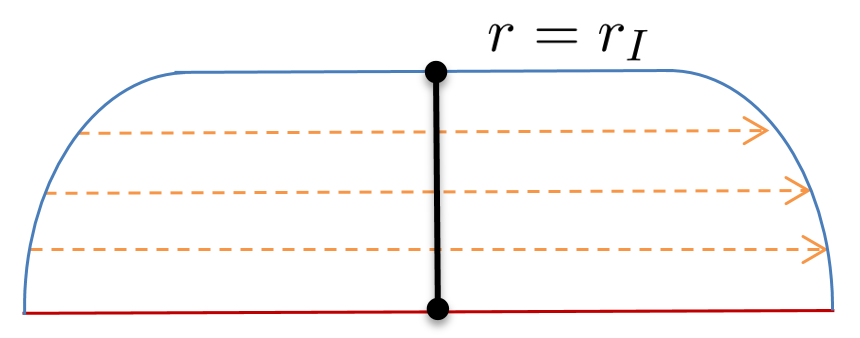}
 \caption{The AdS$_2$ slices in the hyperbolic black hole. The red line is the null modular flow line of the black point on $\tilde{\mathcal{E}}_{\tilde{\mathcal{A}}}$ on the horizon. The orange dashed arrow are the modular flows along the $\tilde{\tau}$ direction. On the left figure the blue line maps to the modular flow line at $r=\infty$, while on the right figure the blue line maps to the modular flow line pushed to $r=r_I$. Note that, this line is not a modular flow line unless we take the limit $r_I\to \infty$.
\label{ads2slices} }
\end{figure}

The bulk and boundary modular flows are just given by $\partial_{\tilde{\tau}}$. It is easy to see that the hyperbolic black hole can be considered as a slicing of the AdS$_2$ slices with fixed $\tilde{x}$. The left figure in Fig.\ref{ads2slices} shows a typical AdS$_2$ slice where the boundary is settled at the $\tilde{r}=\infty$. Since the boundary modular flow is also the bulk modular flow, the trajectory of the boundary modular flow line (the blue line) is itself rather than a two-dimensional slice, thus the concept of modular plane seems not to exist. This subtlety also exist when we look at the original AdS space. The key to understand this subtlety is to push the boundary modular flow line in the original AdS$_3$ from $r=\infty$ to $r=r_I$ ($r_I$ is large but still finite) first, then take the limit $r_I\to \infty$. We argue that the line at $r=r_I$ maps to the blue line in the right figure of Fig.\ref{ads2slices}. Because both of the large $\tilde{r}$ and large $|\tilde{\tau}|$ region in the Rindler bulk map to the large $r$ region in the original AdS$_3$. So the blue line will intersect with the horizon (the red line) at some large $|\tilde{\tau}|$. Also let us look at the original AdS$_3$, the red line maps to the null geodesics emanating normally from the RT surface. These null geodesics form the null boundary of the entanglement wedge and approach (at large $|\tilde{\tau}|$) the two tips of the causal development $\mathcal{D}_{\mathcal{A}}$ at the boundary (at large $r$). 

In this case the blue line have an orbit under the modular flow $\partial_{\tilde{\tau}}$, thus the corresponding modular plane is just the AdS$_{2}$ slice. Then we take the limit $r_I\to\infty$ and the right figure approaches the left figure. In conclusion the modular plane is defined \textit{marginally}.

\begin{figure}[h] 
   \centering
   \includegraphics[width=0.4\textwidth]{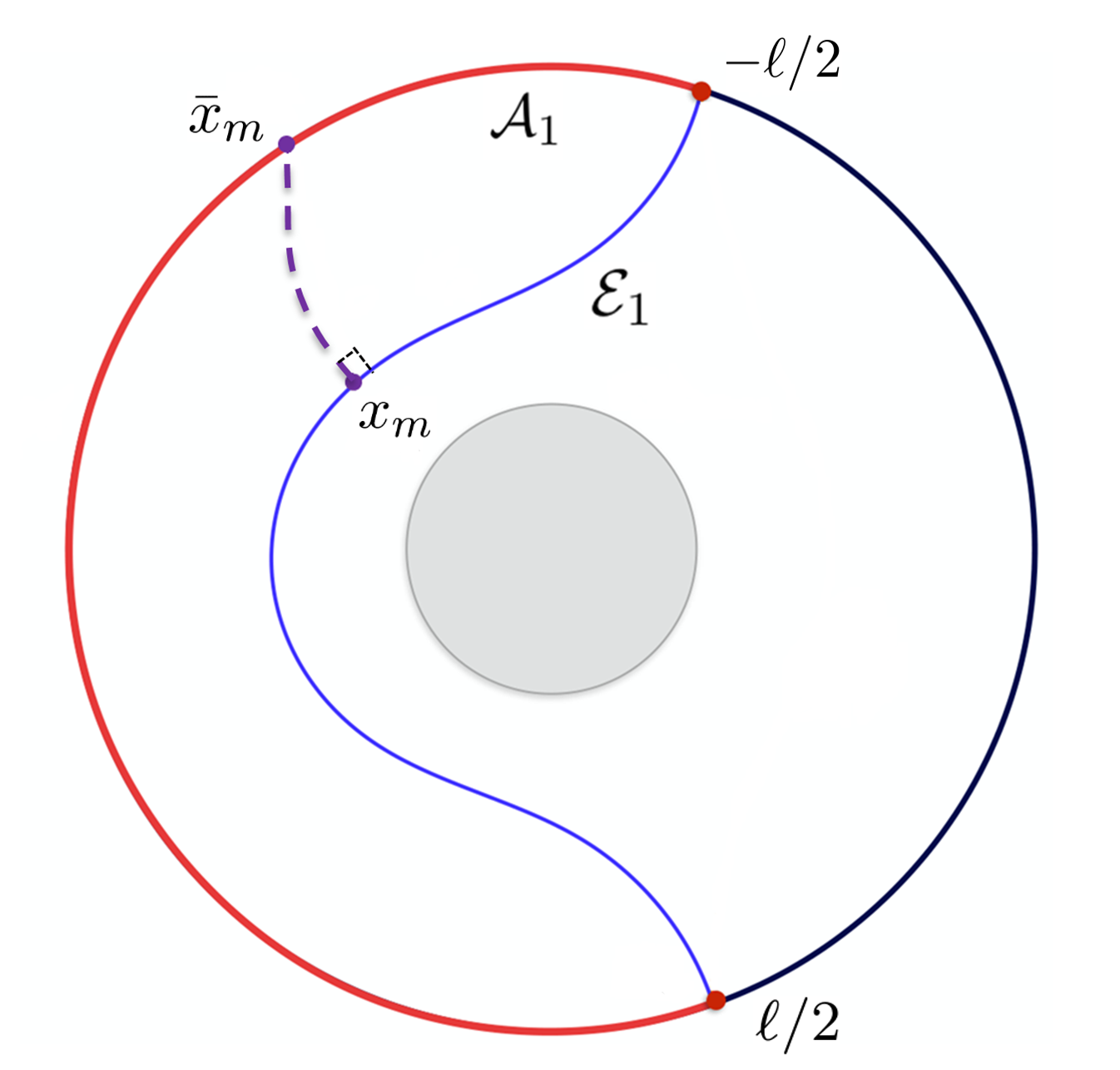} 
 \caption{The gray disk at the center represents the black hole. The dashed line represent the geodesics normal to $\mathcal{E}_{\mathcal{A}}$ which gives the fine correspondence between the points on $\mathcal{A}$ and the points on $\mathcal{E}_{\mathcal{A}}$. 
\label{3} }
\end{figure}

Since the AdS$_2$ slice of the hyperbolic black hole is the modular plane, according to \cite{Wen:2018whg} the points on $\tilde{\mathcal{A}}$ and $\tilde{\mathcal{E}}_{\tilde{\mathcal{A}}}$ that correspond to each other stay in the same AdS$_2$ slice. It is easy to see that each pair of points that correspond to each other can be connected by a geodesic that normal to  the Rindler horizon, which is just the black line in Fig.\ref{ads2slices} along the radius direction. This fine correspondence still holds when we map back to the original BTZ spacetime. In other words the pair of points on $\mathcal{A}$ and $\mathcal{E}_{\mathcal{A}}$ that correspond to each other should be connected by a geodesic normal to $\mathcal{E}_{\mathcal{A}}$. See Fig.\ref{3}.

Consider the following time slice of the BTZ background
\begin{align}
ds^2=\frac{1}{z^2}\left(dx^2+\frac{dz^2}{1-{z^2/z_h^2}}\right)\,,
\end{align}
where $z=z_h$ is the horizon. We consider a static interval $\mathcal{A}$ with length $\ell$ and its middle point settled at $x=0$. In the following we will use the conventions in \cite{Agon:2018lwq} to parameterize the geodesics. Using $(x_m,z_m)$ to parameterize $\mathcal{E}_{\mathcal{A}}$, we have
\begin{align}\label{RTsurface}
\mathcal{E}_{\mathcal{A}}:~~z_m=\sqrt{\frac{z_h^{2}+z_*^{2}}{2}+\frac{z_h^{2}-z_*^{2}}{2}\cosh\frac{2 x_m}{z_h}   }
\end{align}
where $z_*$ is the maximum depth of the RT surface. It is related to $\ell$ by
\begin{align}
\ell=2 z_h \, \text{arctanh}\frac{z_*}{z_h}\,.
\end{align}
Note that when $\ell$ becomes larger than some critical point, the geodesic that compute the entanglement entropy transfer from the connected geodesic to the two component disconnected geodesics \cite{Hubeny:2013gta}. In this paper, we focus on the cases with connected RT surfaces, and leave the disconnected cases for future discussion. 

It is not hard to find out the slope $s_m$ of the vectors normal to $\mathcal{E}_{\mathcal{A}}$ at any point $(x_m,z_m)$, which is given by
\begin{align}\label{sm}
s_m=\pm\frac{z_m}{z_h}\sqrt{\frac{z_h^2-z_m^2}{z_*^2+z_m^2}}\,.
\end{align}
The plus sign correspond to $x_m>0$ while the minus sign correspond to $x_m<0$. Then we consider a static geodesic $\gamma_m$ that passes through $(x_m,z_m)$ with a slope $s_m$. We require the slope to satisfy \eqref{sm} thus $\gamma_m$ is normal to $\mathcal{E}_{\mathcal{A}}$. According to our previous discussion, the partner point of $(x_m,z_m)$ on $\mathcal{A}$ is the point $(\bar{x}_m,0)$ where $\mathcal{A}$ intersect with $\gamma_m$. Following \cite{Agon:2018lwq} $\gamma_m$ can be written as follows:
\begin{align}
\gamma_m:~~z(x)=\sqrt{\mathcal{C}_1+\mathcal{C}_2\cosh\left(\frac{2(x-x_m)}{z_h}\right)+\mathcal{C}_3\sinh\left(\frac{2 (x-x_m)}{z_h}\right)   }
\end{align}
where the coefficients $\mathcal{C}_i$ are given by
\begin{align}\label{Ci}
\mathcal{C}_1=&\frac{(z_h^2+z_m^2)(z_h^2-z_m^2)+s_m^2 z_m^2 z_h^2}{2(z_h^2-z_m^2)}\,,\quad \mathcal{C}_3=s_m z_m z_h\,,
\cr
\mathcal{C}_2=&-\frac{(z_h^2-z_m^2)^2+s_m^2 z_m^2 z_h^2}{2(z_h^2-z_m^2)}\,.
\end{align}
Without loss of generality we focus on the $x_m>0$ part. We find the geodesic $\gamma_m$ intersect with the boundary at the following two points
\begin{align}
x_1=&x_m+\frac{z_h}{2}\log\Big[-\frac{\mathcal{C}_1+\sqrt{\mathcal{C}_1^2-\mathcal{C}_2^2+\mathcal{C}_3^2}}{\mathcal{C}_2+\mathcal{C}_3}\Big]\,,
\cr
x_2=&x_m+\frac{z_h}{2}\log\Big[\frac{-\mathcal{C}_1+\sqrt{\mathcal{C}_1^2-\mathcal{C}_2^2+\mathcal{C}_3^2}}{\mathcal{C}_2+\mathcal{C}_3}\Big]\,.
\end{align}
Only the one with the smaller $x$ coordinate belongs to $\mathcal{A}$.  One can check that $x_2<x_1$ thus $\bar{x}_m=x_2$. Using the definitions \eqref{Ci} we have
\begin{align}\label{bxmxm}
\bar{x}_m=&\frac{1}{2} z_h \log \left(\frac{4 \cosh \left(\frac{2 x_m}{z_h}\right)-4 \cosh \left(\frac{\ell}{z_h}\right)+\cosh \left(\frac{2 \ell}{z_h}\right)-1-N}{8 \left(\sinh ^2\left(\frac{\ell}{2 z_h}\right) \cosh \left(\frac{x_m}{z_h}\right)-\cosh ^2\left(\frac{\ell}{2 z_h}\right) \sinh \left(\frac{x_m}{z_h}\right)\right){}^2}\right)\,,
\end{align}
where 
\begin{align}
N=&4 \sqrt{2} \sinh \left(\frac{\ell}{z_h}\right) \sinh \left(\frac{x_m}{z_h}\right) \sqrt{\cosh \left(\frac{\ell}{z_h}\right)-\cosh \left(\frac{2 x_m}{z_h}\right)}\,.
\end{align}
Furthermore one can check that the above fine correspondence \eqref{bxmxm} is consistent with a much simpler equation
\begin{align}\label{bxmxm2}
\frac{\sinh \left(\frac{\ell/2+ x_m}{z_h }\right) }{\text{sinh}\left(\frac{\ell/2- x_m}{z_h }\right)}=\frac{\sinh ^2\left(\frac{\ell/2+  \bar{x}_m}{2 z_h }\right)}{\sinh ^2\left(\frac{\ell/2-  \bar{x}_m}{2 z_h }\right)}\,.
\end{align}
It is not easy to get \eqref{bxmxm2} by directly simplifying \eqref{bxmxm}. We get \eqref{bxmxm2} by combining the \textit{fine structure analysis} and the PEE \textit{proposal} together, which we will show later.
 
Consider an infinite small segment $d\bar{x}_m$ on $\mathcal{A}$, its partner under the fine correspondence is an infinite small segment on $\mathcal{E}_{\mathcal{A}}$, whose length will give the PEE for $d\bar{x}_m$. More explicitly the contour function $f_{\mathcal{A}}(\bar{x}_m)$ can be read from the following equation
\begin{align}
s_{\mathcal{A}}(d\bar{x}_m)=f_{\mathcal{A}}(\bar{x}_m) d \bar{x}_m=\frac{1}{4G}\sqrt{\frac{1}{z_m^2}\left(dx_m^2+\frac{dz_m^2}{1-\frac{z_m^2}{z_h^2}}\right)}\,.
\end{align}
We can solve $z_m$ as a function of $x_m$ from \eqref{RTsurface} and furthermore plug in the fine correspondence \eqref{bxmxm}. At last we find the entanglement contour function is given by
\begin{align}\label{contourfine}
f_{\mathcal{A}}(x)=\frac{1}{4G}\frac{\coth \left(\frac{\ell+2 x}{4 z_h}\right)+\coth \left(\frac{\ell-2 x}{4 z_h}\right)}{2 z_h}\,,
\end{align}
where we have replaced $\bar{x}_m$ with $x$.

On the other hand, since the entanglement entropies for arbitrary single intervals are known in the dual CFT$_2$ with finite temperature, we can use the PEE \textit{proposal} \eqref{ECproposal} to derive the PEE and furthermore the entanglement contour function based on the \textit{proposal}. The entanglement entropy for any static interval with length $\Delta x$ is given by
\begin{align}\label{SE1}
S_{EE}=\frac{c}{3}\log\left(\frac{\beta}{\pi\epsilon}\sinh \left(\frac{\pi}{\beta} \Delta x \right)\right)\,.
\end{align}
 where $\beta=2\pi z_h$. Now we consider the following partition of $\mathcal{A}=\mathcal{A}_1\cup\mathcal{A}_3$ with 
\begin{align}
\mathcal{A}_1:~-\frac{\ell}{2}\to x\,, \qquad \mathcal{A}_3:~ x\to \frac{\ell}{2}\,.
\end{align}
We apply \eqref{SE1} to the subsets $\mathcal{A}_1$ and $\mathcal{A}_3$ to calculate their entanglement entropies and use the \textit{proposal} to calculate the PEE. Then we get
\begin{align}\label{matching3}
s_{\mathcal{A}}(\mathcal{A}_1)&=\frac{1}{2}\left(S_{\mathcal{A}_1}+S_{\mathcal{A}}-S_{\mathcal{A}_3}\right)
=\frac{c}{6}\log\left(\frac{\beta}{\pi\epsilon}\frac{\sinh \left(\frac{\pi}{\beta} (x+\frac{\ell}{2}) \right)\sinh \left(\frac{\pi}{\beta} \ell \right)}{\sinh \left(\frac{\pi}{\beta} (\frac{\ell}{2}-x) \right)}    \right)\,.
\end{align}
The \textit{proposal} furthermore proposes that the entanglement contour on $\mathcal{A}$ should be, 
\begin{align}\label{contourpee}
f_{\mathcal{A}}(x)=\partial_{x}s_{\mathcal{A}}(\mathcal{A}_1)=\frac{c \pi }{6 \beta }\left(\coth \left(\frac{\pi (\ell/2+x)}{ \beta }\right)+\coth \left(\frac{\pi(\ell/2-x)}{ \beta }\right)\right)\,.
\end{align} 
Since $c=\frac{3}{2G}$ and $\beta=2\pi z_h$, the above result exactly matches with our previous contour function \eqref{contourfine} we got from the \textit{fine structure analysis}.

\subsection{Correspondence between bulk geodesic chords and boundary PEE}
The fine correspondence between the points on the interval $\mathcal{A}$ and its corresponding RT surface $\mathcal{E}_{\mathcal{A}}$ can be extend to the correspondence between any geodesic chords $\mathcal{E}_i$ to the PEE of a subset $\mathcal{A}_i$, and furthermore to a linear combination of subset entanglement entropies according to our \textit{proposal} \eqref{ECproposal}. For any geodesic chord $\mathcal{E}_i$ we extend it to a whole geodesic $\mathcal{E}_{\mathcal{A}}$ anchored on the boundary which correspond to a boundary interval $\mathcal{A}$ according to the RT proposal. Then $\mathcal{A}_i$ is the subset of $\mathcal{A}$ that correspond to $\mathcal{E}_i$ according to the fine correspondence. In other words, we have
\begin{align}\label{pechord}
 s_{\mathcal{A}}(\mathcal{A}_i)=\frac{Length\left(\mathcal{E}_i\right)}{4G}\,.
 \end{align} 
Also the similar construction is conducted for WCFT in \cite{Wen:2018mev} in the context of AdS$_3$/WCFT correspondence \cite{Detournay:2012pc,Compere:2013bya}.

Now we use the PEE \textit{proposal} to determine the fine correspondence in a much simpler way and derive the simpler relation \eqref{bxmxm2}. Since all the subset entanglement entropies are known, we can use our \textit{proposal} \eqref{ECproposal} to calculate $s_{\mathcal{A}}(\mathcal{A}_i)$ and then impose the matching condition \eqref{pechord} for the bulk geodesic chords. The strategy is to consider a pair of points $\mathcal{A}(\bar{x}_m)$ and $\mathcal{E}_{\mathcal{A}}(x_m)$ on $\mathcal{A}$ and $\mathcal{E}_{\mathcal{A}}$ that lead to the partitions $\mathcal{A}=\mathcal{A}_1\cup \mathcal{A}_3\,,~ \mathcal{E}_{\mathcal{A}}=\mathcal{E}_1\cup\mathcal{E}_3\,.$
Then these two points are correspond to each other when they satisfy the matching condition \eqref{pechord}. The right hand side of \eqref{pechord} can be easily calculated by integrating the length of the geodesic chords. Integrating $ds$ along the RT surface from $x=-\frac{\ell}{2}+\delta_x$ to $x=x_m$ we get
\begin{align}
\frac{Area(\mathcal{E}_1)}{4G}=\frac{c}{12}\log \Big[\frac{\beta}{ 2\pi \delta_x}\sinh \left(\frac{2\pi}{\beta}(\ell/2+x_m)\right)  \sinh \left(\frac{2\pi \ell}{\beta}\right)  \text{csch}\left(\frac{2\pi}{\beta}(\ell/2-x_m)\right)\Big]
\end{align}
The length of $\mathcal{E}_{1}$ is infinite, so we use $(-\frac{\ell}{2}+\delta_x, \delta_z)$ to denote the point where we cut off the RT surface. Since $\delta_z$ is related to the UV cutoff $\epsilon$ by $\delta_z=\epsilon$, we have $\delta_x=\frac{\pi   }{\beta } \coth\left(\frac{\pi  \ell}{\beta }\right)\epsilon ^2$, such that
\begin{align}\label{matching2}
\frac{Area(\mathcal{E}_1)}{4G}=\frac{c}{12}\log \Big[&\frac{\beta^2}{\pi^2\epsilon^2}\sinh \left(\frac{2 \pi  (\ell/2+x_m)}{\beta }\right)\sinh ^2\left(\frac{\pi  \ell}{\beta }\right)
 \text{csch}\left(\frac{2 \pi  ((\ell/2-x_m))}{\beta }\right)\Big]
\end{align}
On the other hand the PEE $S_{\mathcal{A}}(\mathcal{A}_1)$ can be calculated using the PEE \textit{proposal}. The result is given by \eqref{matching3} with the $x$ replaced with $\bar{x}_m$. Then according to \eqref{pechord} we match \eqref{matching2} with \eqref{matching3} to get
\begin{align}\label{bxmxm3}
\frac{\sinh \left(\frac{2 \pi  (\ell/2+x_m)}{\beta }\right) }{\text{sinh}\left(\frac{2 \pi  ((\ell/2-x_m)}{\beta }\right)}=\frac{\sinh ^2\left(\frac{\pi  (\ell/2+\bar{x}_m)}{\beta }\right)}{\sinh ^2\left(\frac{\pi  (\ell/2-\bar{x}_m)}{\beta }\right)}\,.
\end{align}
This relation is just \eqref{bxmxm2} which coincide with the fine correspondence \eqref{bxmxm} we get from the \textit{fine structure analysis} in the entanglement wedge.

For an arbitrary static geodesic chord $\mathcal{E}_2$ settled on a static geodesic, which is the RT surface of a boundary interval with length $\ell$ and its middle point settled at $x=0$. Assuming the $x$ coordinate of the end points of $\mathcal{E}_2$ are given by $x_2$ and $x_1$ with $x_2>x_1$, then the length of  $\mathcal{E}_2$ is given by the PEE $s_{\mathcal{A}}(\mathcal{A}_2)$ with the subsets given by
\begin{align}
\mathcal{A}_1:-\frac{\ell}{2}\leq x \leq \bar{x}_1\,,\qquad\mathcal{A}_2:\bar{x}_1\leq x \leq \bar{x}_2\,,\qquad \mathcal{A}_3:\bar{x}_2\leq x \leq \frac{\ell}{2}\,,
\end{align}
where $\bar{x}_1$ and $\bar{x}_2$ are determined by $x_1$ and $x_2$ via \eqref{bxmxm3}. More interestingly the length of any geodesic chords can be written as a linear combination of entanglement entropies of certain boundary intervals,
\begin{align}
\frac{Area\left(\mathcal{E}_2\right)}{4G}=s_{\mathcal{A}}(\mathcal{A}_2)=\frac{1}{2}\left(S_{\mathcal{A}_1\cup\mathcal{A}_2}+S_{\mathcal{A}_2\cup\mathcal{A}_3}-S_{\mathcal{A}_1}-S_{\mathcal{A}_3}\right)
\end{align}
This is a finer version of the RT formula and plays a similar role as the \textit{kinematic space} formalism \cite{Balasubramanian:2013lsa,Czech:2014ppa,Czech:2015qta,Czech:2015kbp}. Some explicit comparison between the \textit{kinematic space} picture and the \textit{fine structure analysis} picture can be found in \cite{Abt:2018ywl,Kudler-Flam:2019oru}.

\section{Generating local modular flows from entanglement entropy}
The entanglement contour contains much more information than entanglement entropy alone. Then it is interesting to ask whether the entanglement contour proposed by \eqref{ECproposal} contains all the information needed to reconstruct the reduced density matrix (or modular Hamiltonian). In this section we show that, for the cases where the modular Hamiltonian is local thus generate a local modular (geometric) flow, the modular flow can be extracted from the PEE via our \textit{proposal}.  This involves an interesting property of the PEE observed in \cite{Wen:2018whg}. More explicitly for an arbitrary subset $\alpha$ of an interval $\mathcal{A}$, we have
\begin{align}\label{minvariant}
 s_{\mathcal{A}}(\alpha)=s_{\mathcal{A}'}(\alpha')\,.
 \end{align}
Here $\mathcal{A}'$ could be any spacelike interval that has the same causal development as $\mathcal{A}$, i.e. $\mathcal{D}_{\mathcal{A}'}=\mathcal{D}_{\mathcal{A}}$. And $\alpha'$ is the subset of $\mathcal{A}'$ which intersect with the same class of modular flow lines (or modular planes) as $\alpha$. In other words the orbits of two end points of $\alpha$ under the modular flow is a pair of modular flow lines. Any spacelike sub-interval $\alpha'$ whose end points respectively settle on this pair of modular flow lines will satisfy \eqref{minvariant}. This property is obvious in the \textit{fine structure analysis} of the entanglement wedge \cite{Wen:2018whg} since the cyclic gluing of both $\alpha$ and $\alpha'$ turn on the replica story on the same class of modular planes thus turn on contribution from the same bulk geodesic chord. We see no obstacle to extend this property to covariant intervals and furthermore to other holographic cases with local modular flow. The more interesting question is whether this property can be extend to the non-holographic cases.

Previously the only way to solve the local modular flow relies on the Rindler transformations, which are symmetry transformations that map the causal development of an interval to a Rindler space with infinitely far away boundaries. The translation along the Rindler time in the Rindler space then maps to the modular flow in the causal development under the inverse Rindler transformations. The Rindler transformations for static intervals in the vacuum state of CFT$_2$ and static spheres in the vacuum state of higher dimensional CFTs are firstly constructed in \cite{Casini:2011kv}. Then in \cite{Song:2016gtd,Jiang:2017ecm} a prescription to construct Rindler transformations in 2-dimensional theories using the global symmetries was proposed. Following this prescription the Rindler transformations and also the modular flows for any intervals (beyond the static case) have been constructed for CFT$_2$ in the vacuum state \cite{Song:2016gtd}, warped CFT \cite{Song:2016gtd,Castro:2015csg} and field theories invariant under the BMS$_3$ transformations (BMSFTs) \cite{Jiang:2017ecm}. Though the prescription is quite successful in 2-dimensional theories, it is not easy to carry out. Even for the case of CFT$_2$ in a thermal state, the construction of the Rindler transformations has not been carried out yet. Nevertheless the modular flow in this case is carried out in \cite{Jiang:2019qvd} using other tricks.

\begin{figure}[h] 
   \centering
   \includegraphics[width=0.45\textwidth]{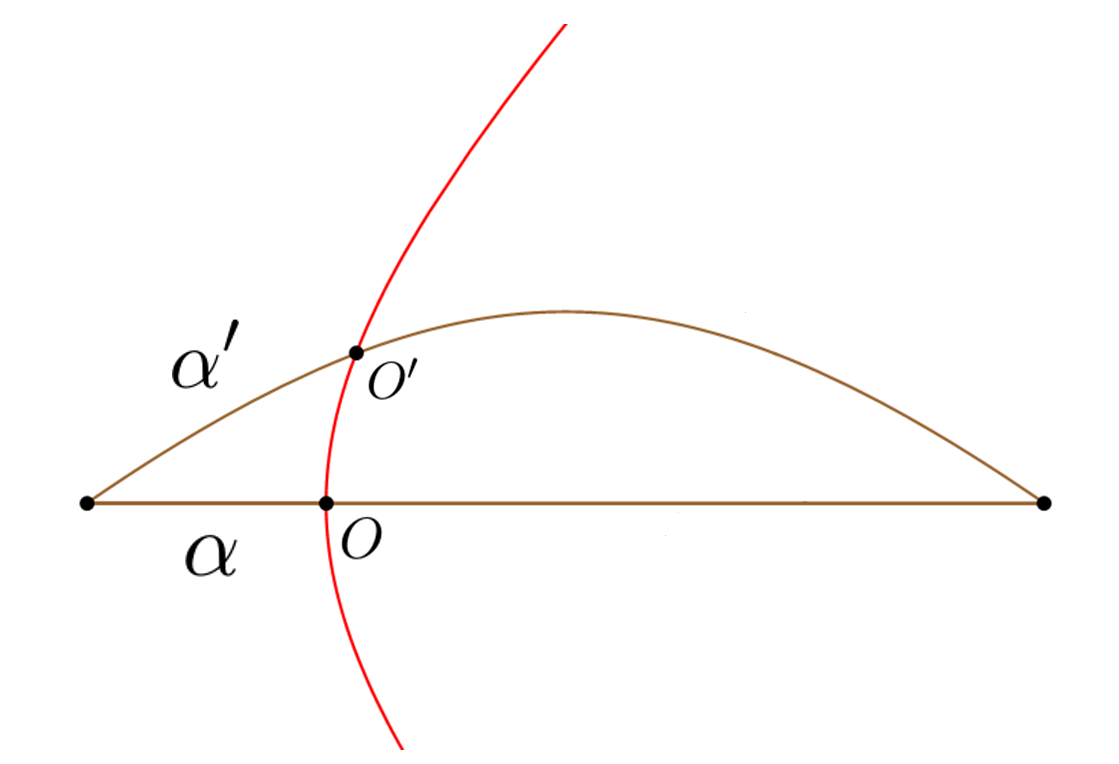} 
 \caption{The red line consist of all the points $O'$ that keeps $s_{\mathcal{A}'}(\alpha')=s_{\mathcal{A}}(\alpha)$. The red line is the modular flow line according to the property \eqref{minvariant}.
\label{5} }
\end{figure}

In this section, we propose a much simpler prescription to generate the local modular flow for single intervals in 2-dimensional field theories from entanglement entropy. Provided the entanglement entropy for any intervals are known, the PEE $s_{\mathcal{A}'}(\alpha')$ for any sub-interval $\alpha'$ in the causal development $\mathcal{D}_{\mathcal{A}}$ can be calculated via the PEE \textit{proposal} \eqref{ECproposal}. Here we need combine the property \eqref{minvariant} with the PEE \textit{proposal}. It will be more convenient to consider the partition $\mathcal{A}=\alpha\cup\alpha_R$ with $\alpha_L$ vanished thus the left end point of $\alpha$ coincide with the left end point of $\mathcal{A}$. We denote the other end point of $\alpha$ as $O$. We only consider $\mathcal{A}'$ that shares end points with $\mathcal{A}$. The requirement $\mathcal{D}_{\mathcal{A}}=\mathcal{D}_{\mathcal{A}'}$ is satisfied by this choice of $\mathcal{A}'$. We consider a similar partition $\mathcal{A}'=\alpha'\cup\alpha'_R$, where $\alpha'$ also shares the left end point with $\alpha$. We denote the right end point $\alpha'$ as $O'$ (see Fig.\ref{5}). Then the modular flow line that passes through $O$ should consist of all the points $O'$ that satisfy \eqref{minvariant}. In the following we will explicitly use this strategy to generate the modular flows for intervals in CFT$_2$, warped CFT and BMSFT.

\subsection*{Modular flows for CFT$_2$ in a vacuum state or thermal state}
In order to describe the modular flow and thermal circle of the state, we need to embed the theory in a coordinate system. It is convenient to consider the CFT$_2$ duals to the following BTZ background in light-cone coordinates 
\begin{align}\label{mBTZ}
ds^2=\frac{dr^2}{4(r^2-T_u^2 T_v^2)}+2 r du dv+T_u^2 du^2+T_v^2 dv^2\,,
\end{align}
where $u$ and $v$ are the light-cone coordinates, and $T_u~(T_v)$ is the left (right) moving temperatures. We consider an arbitrary interval with the following end points
\begin{align}
\mathcal{A}: ~(-\frac{l_u}{2},-\frac{l_v}{2})\to (\frac{l_u}{2},\frac{l_v}{2})\,.
\end{align}

When $T_u=T_v=0$ the metric \eqref{mBTZ} describes the Poincar\'e AdS$_3$ which duals to the vacuum state. The modular flows in the dual CFT is given by \cite{Wen:2018mev}
\begin{align}
k_t=\frac{\pi}{2}\left(\frac{4 u^2}{l_u}-l_u\right)\partial_u-\frac{\pi}{2}\left( \frac{4 v^2}{l_v}-l_v\right)\partial_v\,.
\end{align}
Integration along the modular flow will give us the modular flow lines
\begin{align}\label{mfline1}
v(u)=-\frac{l_v}{2}  \tanh \left(\tanh ^{-1}\left(\frac{2 u}{l_u}\right)-2 c_0 l_v\right)\,.
\end{align}
where $c_0$ is the integration constant that characterizes different modular flow lines.

Then we use our prescription to derive the modular flow lines \eqref{mfline1}. The entanglement entropy for any interval is given by
\begin{align}
S_{EE}=\frac{c}{6}\log\frac{\Delta u\Delta v}{\epsilon_u\epsilon_v}\,,
\end{align}
where $\Delta u$ and $\Delta v$ are the coordinate differences between the two end points and $\epsilon_u$ ($\epsilon_v$) is the cutoff on the left (right) moving sector.  Assuming all the $O'$ that satisfy \eqref{minvariant} form a line described by the function $(u,v(u))$. Then according to the \textit{proposal} \eqref{ECproposal} we have
\begin{align}
s_\mathcal{A'}(\alpha')=\frac{c}{12}\left(\log\frac{(u+l_u/2)(v(u)+l_v/2)l_u l_v}{\epsilon_u\epsilon_v (l_u/2-u)(l_v/2-v(u))}\right)\,.
\end{align}
According to \eqref{minvariant}, the PEE $s_\mathcal{A'}(\alpha')$ should be a constant along the line $(u,v(u))$. So we should have 
\begin{align}
\partial_u s_\mathcal{A'}(\alpha')=0,
\end{align}
 which reduces in the following equation
\begin{align}
 l_v \left(\left(l_u^2-4 u^2\right) v'(u)+l_u l_v\right)-4  l_u v(u)^2=0\,.
 \end{align} 
 The solution is given by
 \begin{align}
 v(u)= -\frac{l_v}{2}  \tanh \left(\tanh ^{-1}\left(\frac{2 u}{l_u}\right)-2 c_1 l_v\right)
 \end{align}
 where $c_1$ is the integration constant. Choosing $c_1=c_0$, the above solution is exactly \eqref{mfline1} thus reproduces the modular flow lines.
 
For the cases with a thermal circle, the left and right moving temperatures are usually nonzero. The modular flow is obtained in \cite{Jiang:2019qvd}, which is given by
\begin{align}
k_t=&\frac{\pi  \sinh \left(l_v T_v\right) \left(\cosh \left(l_u T_u\right)-\cosh \left(2 u T_u\right)\right)}{l_u l_v T_u^2 T_v}\partial_u
\cr
&+\frac{\pi  \sinh \left(l_u T_u\right) \left(\cosh \left(2 v T_v\right)-\cosh \left(l_v T_v\right)\right)}{l_u l_v T_u T_v^2}\partial_v\,.
\end{align}
The corresponding modular flow lines are given by
\begin{align}\label{mfline2}
v(u)=-\frac{\tanh ^{-1}\Big[\tanh \left(\frac{l_v T_v}{2}\right) \tanh \left(\tanh ^{-1}[\tanh \left(u T_u\right) \coth \left(\frac{l_u T_u}{2}\right)]-c_0 T_v \sinh \left(l_v T_v\right)\right)\Big]}{T_v}\,,
\end{align}
where $c_0$ is again an integration constant.

Now we reproduce the modular flow lines \eqref{mfline2} using the PEE \textit{proposal} and \eqref{minvariant}. The entanglement entropy for an arbitrary interval is given by
\begin{align}
S_{EE}=\frac{c}{6}\log\left(\frac{\sinh(\Delta u T_u)\sinh(\Delta v T_v)}{T_u T_v \epsilon_u\epsilon_v}\right)\,,
\end{align}
hence, according to \eqref{ECproposal} we have
\begin{align}
s_\mathcal{A'}(\alpha')=\frac{c}{12} \log \Big[\frac{\sinh \left(l_u T_u\right) \sinh \left(l_v T_v\right) \sinh \left[\left(\frac{l_u}{2}+u\right) T_u\right] \sinh \left[\left(\frac{l_v}{2}+v(u)\right)T_v \right]}{\sinh \left[\left(\frac{l_u}{2}-u\right) T_u\right] \sinh \left[T_v \left(\frac{l_v}{2}-v(u)\right)\right]T_u T_v \epsilon_u\epsilon_v}\Big]
\end{align}
The equation $\partial_u s_\mathcal{A'}(\alpha')=0$ then reduces to
\begin{align}
\frac{T_v   \sinh \left(l_v T_v\right)\sinh \left[\left(\frac{l_u}{2}+u\right) T_u\right]}{\text{sinh}\left[T_v \left(\frac{l_v}{2}-v(u)\right)\right]}v'(u)+\frac{T_u \sinh \left(l_u T_u\right) \sinh \left[ \left(\frac{l_v}{2}+v(u)\right)T_v\right]}{\text{sinh}\left[\left(\frac{l_u}{2}-u\right) T_u\right]}=0
\end{align}
One can check that the solution of the above equation is exactly \eqref{mfline2} which describes the modular flow lines.

\subsection*{Modular flows for WCFT}
The AdS$_3$ \eqref{mBTZ} with a Dirichlet-Neumann type of boundary conditions \cite{Compere:2013bya} duals to a WCFT\cite{Detournay:2012pc}. Here we focus on the case with $T_u=0,\,T_v=1$. The modular flow in this case was carried out in \cite{Wen:2018mev} using the Rindler method,
\begin{align}
k_t=\frac{\pi}{2}\left(\frac{4 u^2}{l_u}-l_u\right)\partial_u+\pi\partial_v\,.
\end{align}
The corresponding modular flow lines are given by
\begin{align}\label{mfline3}
v(u)=c_0-\tanh ^{-1}\left(\frac{2 u}{l_u}\right)\,,
\end{align}
where $c_0$ is an integration constant.

On the other hand, the entanglement entropy for an arbitrary interval in WCFT is given by
\begin{align}
S_{EE}=\frac{c}{6}\left(\Delta v+\log\frac{\Delta u}{\epsilon_u}\right)
\end{align}
Then the PEE \textit{proposal} gives
\begin{align}
s_\mathcal{A'}(\alpha')=\frac{c}{12}  \left(\log \left(\frac{l_u \left({l_u/2}+u\right)}{({l_u/2}-u)\epsilon_u}\right)+l_v+2 v(u)\right)\,,
\end{align}
and the condition $\partial_u s_\mathcal{A'}(\alpha')=0$  reduces to
\begin{align}
\frac{c}{6}  \left(\frac{2 l_u}{l_u^2-4 u^2}+v'(u)\right)=0\,.
\end{align}
The solution of the above condition is exactly the modular flow lines \eqref{mfline3}.

For WCFT in general temperatures the entanglement entropy becomes
\begin{align}
S_{EE}=\frac{c}{6}\left(\Delta v T_v+\log\frac{\sinh(\Delta u T_u)}{\epsilon_u T_u}\right)\,.
\end{align}
Following our prescription we can easily get the corresponding modular flow lines
\begin{align}
v(u)=c_0+\frac{\log \left[\sinh \left(\frac{l_u T_u}{2}-u T_u\right) \text{csch}\left(\frac{l_u T_u}{2}+u T_u\right)\right]}{2 T_v}\,.
\end{align}

\subsection*{Modular flow for BMSFTs}
The third example is the reconstruction of the modular flow lines of the BMSFTs, which is conjectured to be the field theory dual of the 3-dimensional flat space \cite{Barnich:2010eb,Bagchi:2010eg,Bagchi:2012cy}
\begin{align}\label{fbtz}
ds^2= M du^2-2 dudr+ J du d\phi+r^2 d\phi^2.
\end{align}
Here $M$ and $J$ denote the mass and angular momentum of the space. Depending on the values of $M$ and $J$, the above solutions are usually classified into three types, the Global Minkowski, Null-orbifold and the Flat Space Cosmological solutions. The asymptotic boundary at $r=\infty$ is a null surface with a spacelike coordinate $\phi$ and null coordinate $u$. Here we only consider the case of Null-orbifold with $M=J=0$, which duals to the vacuum state of the BMSFT that lives on a null plane. When the gravity theory is Einstein gravity, then one of the central charge of the dual BMSFT vanishes, i.e. $C_L=0$. We consider a general interval with the following end points 
\begin{align}
\mathcal{A}:~(-\frac{l_\phi}{2},-\frac{l_u}{2})\to (\frac{l_\phi}{2},\frac{l_u}{2}).
\end{align}
The corresponding modular flow is carried out in \cite{Jiang:2017ecm} using the Rindler method,
\begin{align}
k_t=-{\pi\over 2  l_\phi}   \left(( l_\phi ^2-4 \phi^2) \partial_\phi \,+   ( l_ul_\phi +4 {l_u\over l_\phi }\phi^2-8 u \phi ) \partial_u\right).
\end{align}
Integrating along the modular flow $k_t$ we get the modular flow lines 
\begin{align}\label{mfline4}
u=c_0( 1-\frac{4  \phi^2}{l_\phi^2})+\frac{l_u\phi  }{l_\phi}\,,
\end{align}
where $c_0$ is the integration constant that characterizes all the modular flow lines.

Then we try to reproduce the modular flow lines \eqref{mfline4}. In this case the entanglement entropy for an arbitrary interval is given by \cite{Bagchi:2014iea,Basu:2015evh,Jiang:2017ecm,Hijano:2017eii}
\begin{align}\label{sbms}
S_{\mathcal{A}}=\frac{c_M}{6}\frac{\Delta u}{\Delta\phi}
\end{align}
where $c_M$ is the central charge. Note that, we have set the other central charge $c_L=0$ thus the gravity theory is just the Einstein gravity. Assuming the modular flow line is parametrized by the function $(u(\phi),\phi)$. Then the PEE \textit{proposal} gives
\begin{align}\label{sa1'}
s_{\mathcal{A}'}&(\alpha')=\frac{c_M}{12}\left(\frac{l_u \left(l_\phi^2-4 l_\phi \phi -4 \phi ^2\right)+4 l_\phi^2 u(\phi )}{l_\phi^3-4 l_\phi \phi ^2}\right)\,.
\end{align}
The condition $\partial_\phi s_{\mathcal{A}'}(\alpha')=0$ then gives the following solution
\begin{align}
u(\phi )= c_1 \left(l_\phi^2-4 \phi ^2\right)+\frac{l_u \phi }{l_\phi}
\end{align}
where $c_1$ is an integration constant. This reproduces the modular flow lines \eqref{mfline4} by a redefinition of the constant $c_0= c_1 l_\phi^2$.

The extension of our prescription to the case of Global Minkowski or the Flat Space Cosmological solutions is straightforward.

\section{Discussion}
\begin{figure}[h] 
   \centering
    \includegraphics[width=0.45
   \textwidth]{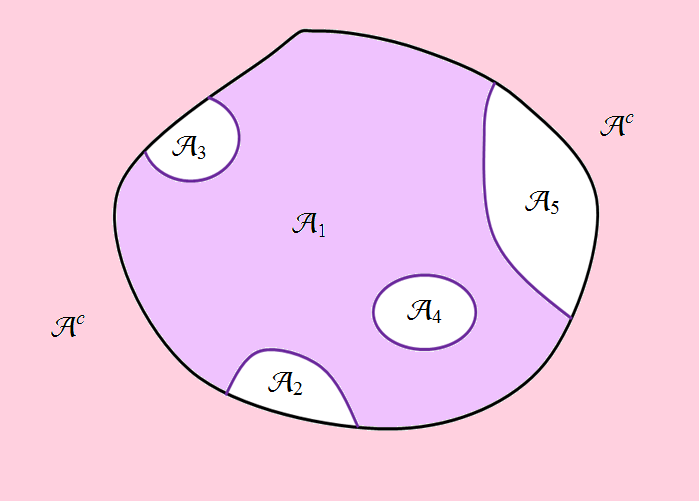}
 \caption{An example of a more general partition in higher dimensions. The subset $\mathcal{A}_1$ divide the region $\mathcal{A}=\mathcal{A}_1\cup\mathcal{A}_2\cup\mathcal{A}_3\cup\mathcal{A}_4\cup\mathcal{A}_5$ into five subsets. 
\label{partition2} }
\end{figure}

Since we only used the construction of the proposal \eqref{ECproposal} and the general inequalities satisfied by entanglement entropies, the PEE \textit{proposal} should satisfy the above 6 properties in general two-dimensional theories.
%According to our proposal, for one-dimensional $\mathcal{A}$ we always have
%\begin{align}
%s_{\mathcal{A}}(\mathcal{A}^{a}_{2})&=\frac{1}{2}\left(S_{\mathcal{A}_1\cup\mathcal{A}^{a}_{2}}+S_{\mathcal{A}^{}_{2}\cup\mathcal{A}^{}_{3}}-S_{\mathcal{A}^{}_{1}}-S_{\mathcal{A}^{b}_{2}\cup\mathcal{A}^{}_{3}} \right)\,,
%\cr
%s_{\mathcal{A}}(\mathcal{A}^{b}_{2})&=\frac{1}{2}\left(S_{\mathcal{A}_1\cup\mathcal{A}^{}_{2}}+S_{\mathcal{A}^{b}_{2}\cup\mathcal{A}^{}_{3}}-S_{\mathcal{A}^{}_{1}\cup\mathcal{A}^{a}_{2}}-S_{\mathcal{A}^{}_{3}} \right)\,,
%\end{align}
%thus $s_{\mathcal{A}}(\mathcal{A}^{}_{2})=s_{\mathcal{A}}(\mathcal{A}^{a}_{2})+s_{\mathcal{A}}(\mathcal{A}^{b}_{2})$.
One important underlying ingredient of our proposal is that, no matter how we do the partition the subsets that forms $\mathcal{A}$ are always in sequential \textit{order}. This is crucial for the PEE \eqref{ECproposal} to be additive. However, in general the partitions in higher dimensions are more complicated and there is no natural way to define the order for a general partition, For example, see the case in Fig.\ref{partition2}. In these cases we need to impose further constrains on the entanglement entropies to make the PEE additive. 

\begin{figure}[h] 
   \centering
    \includegraphics[width=0.30 \textwidth]{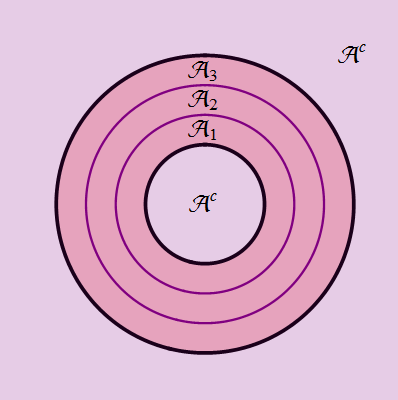}\qquad \qquad  ~~
        \includegraphics[width=0.308 \textwidth]{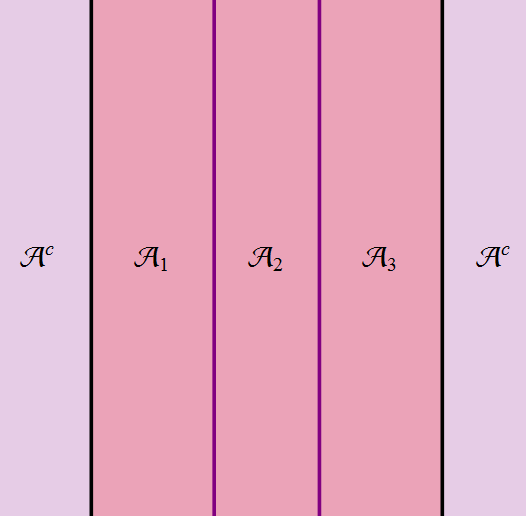}
 \caption{Quasi-one-dimensional configurations with rotation and translation symmetries. Here $\mathcal{A}=\mathcal{A}_1\cup\mathcal{A}_2\cup\mathcal{A}_3$
\label{higherd} }
\end{figure}

Since the entanglement contour function respect the symmetries, for highly symmetric configurations\footnote{Here configuration including the theory, the region $\mathcal{A}$ and its partition.} the contour functions may only depend on one coordinate.  For example consider $\mathcal{A}$ to be a spherical shell in a theory that is rotational symmetric with respect to the center of the shell, the rotation symmetry implies that the contour function only depend on the radius coordinate. Note that the contour function does not depend on how we do the partition, so it is enough to only consider partitions that respect the symmetry, i.e. with all the subsets being concentric spherical shells (see the left figure of Fig.\ref{higherd}). Since a natural \textit{order} can be defined along the radius coordinate, the PEE \textit{proposal} in these configurations should also satisfy all the six requirements \cite{Vidal}. Similarly we can consider configurations with translation symmetries and with the partition shown in the right figure of Fig.\ref{higherd}. We call these kind of configurations the \textit{quasi-one-dimensional configurations}. A more explicit discussion on the entanglement contour for quasi-one dimensional cases in higher dimensions can be found in \cite{Han:2019scu}.  

For holographic theories the \textit{fine structure analysis} indicates that the PEE is invariant under the modular flow of $\mathcal{A}$ in the sense of \eqref{minvariant}. However this is not obvious at all from the field theory side. As we have seen using the PEE \textit{proposal} and the property \eqref{minvariant} we can generate the right modular flows in a much easier way. These are highly non-trivial results and gives strong support for the PEE \textit{proposal}. It will be interesting to test the property \eqref{minvariant} in more general theories that are not holographic. 

Combine our proposal with the holographic picture, we get the correspondence between bulk geodesic chords and boundary PEEs, which can be considered as a finer version of the RT formula or its analogues \cite{Song:2016pwx,Song:2016gtd,Jiang:2017ecm,Wen:2018mev} in holographic theories beyond AdS/CFT. One can consult \cite{Abt:2018ywl} for an interesting application of this correspondence to interpret the bulk volume with boundary entanglement entropies. Since the minimal entanglement wedge cross sections are geodesic chords, the PEE can also be closely related to the quantities that are assumed to be dual to the minimal cross section, which include the entanglement of purification \cite{Takayanagi:2017knl,Nguyen:2017yqw} and the logarithmic negativity \cite{Kudler-Flam:2018qjo}.

On the other way around, if we can calculate the entanglement contour using the fine correspondence in holography \cite{Wen:2018whg,Kudler-Flam:2019oru} or other methods (for example the \textit{Gaussian formula} \cite{Botero,Vidal,PhysRevB.92.115129,Coser:2017dtb,Tonni:2017jom}) for lattice models , we can calculate the entanglement entropies of the subsets with less symmetries via the PEE \textit{proposal} \cite{Han:2019scu}. 

As we have noted the study of entanglement contour is still at an initial stage. Many foundational questions remain to be answered. For example, in general can the entanglement contour be uniquely determined? How can we generalize the PEE \textit{proposal} to disconnected regions? Does the entanglement contour exist in cases with non-local modular Hamiltonian? If it exists then how can we extend the above discussions to the non-local cases\footnote{A possible entry for this problem is to follow the extension \cite{Faulkner:2017vdd} of the reconstruction of bulk operators from the local to non-local cases.}? We leave all these questions for future investigation.

\textit{Note}: During the accomplishment of this paper, \cite{Kudler-Flam:2019oru} appears which has some overlap with our paper. The discussion on uniqueness of the entanglement contour and the justification of the PEE \textit{proposal} in Poincar\'e invariant field theories can be found in a later paper \cite{Wen:2019iyq}.

\section*{Acknowledgement}
We would like to thank Jonah Kudler-Flam, Muxin Han, and Gang Yang for helpful discussions. We would also like to thank Chang-pu Sun and Chuan-jie Zhu for support. This work is supported by a NSFC Grant No.11805109 and the start up funding from the Southeast University.

\bibliographystyle{JHEP}
 \bibliography{lmbib}

\providecommand{\href}[2]{#2}\begingroup\raggedright\begin{thebibliography}{10}

\bibitem{Wen:2018whg}
Q.~Wen, \emph{{Fine structure in holographic entanglement and entanglement
  contour}}, \href{http://dx.doi.org/10.1103/PhysRevD.98.106004}{\emph{Phys.
  Rev.} {\bf D98} (2018) 106004}, [\href{https://arxiv.org/abs/1803.05552}{{\tt
  1803.05552}}].

\bibitem{PhysRevLett.96.110405}
M.~Levin and X.-G. Wen, \emph{Detecting topological order in a ground state
  wave function},
  \href{http://dx.doi.org/10.1103/PhysRevLett.96.110405}{\emph{Phys. Rev.
  Lett.} {\bf 96} (Mar, 2006) 110405}.

\bibitem{Kitaev:2005dm}
A.~Kitaev and J.~Preskill, \emph{{Topological entanglement entropy}},
  \href{http://dx.doi.org/10.1103/PhysRevLett.96.110404}{\emph{Phys. Rev.
  Lett.} {\bf 96} (2006) 110404},
  [\href{https://arxiv.org/abs/hep-th/0510092}{{\tt hep-th/0510092}}].

\bibitem{Calabrese:2004eu}
P.~Calabrese and J.~L. Cardy, \emph{{Entanglement entropy and quantum field
  theory}}, \href{http://dx.doi.org/10.1088/1742-5468/2004/06/P06002}{\emph{J.
  Stat. Mech.} {\bf 0406} (2004) P06002},
  [\href{https://arxiv.org/abs/hep-th/0405152}{{\tt hep-th/0405152}}].

\bibitem{Calabrese:2005zw}
P.~Calabrese and J.~L. Cardy, \emph{{Entanglement entropy and quantum field
  theory: A Non-technical introduction}},
  \href{http://dx.doi.org/10.1142/S021974990600192X}{\emph{Int. J. Quant. Inf.}
  {\bf 4} (2006) 429}, [\href{https://arxiv.org/abs/quant-ph/0505193}{{\tt
  quant-ph/0505193}}].

\bibitem{Hsu:2008af}
B.~Hsu, M.~Mulligan, E.~Fradkin and E.-A. Kim, \emph{{Universal entanglement
  entropy in 2D conformal quantum critical points}},
  \href{http://dx.doi.org/10.1103/PhysRevB.79.115421}{\emph{Phys. Rev.} {\bf
  B79} (2009) 115421}, [\href{https://arxiv.org/abs/0812.0203}{{\tt
  0812.0203}}].

\bibitem{Maldacena:1997re}
J.~M. Maldacena, \emph{{The Large N limit of superconformal field theories and
  supergravity}}, \href{http://dx.doi.org/10.1023/A:1026654312961}{\emph{Int.
  J. Theor. Phys.} {\bf 38} (1999) 1113--1133},
  [\href{https://arxiv.org/abs/hep-th/9711200}{{\tt hep-th/9711200}}].

\bibitem{Witten:1998qj}
E.~Witten, \emph{{Anti-de Sitter space and holography}}, {\emph{Adv. Theor.
  Math. Phys.} {\bf 2} (1998) 253--291},
  [\href{https://arxiv.org/abs/hep-th/9802150}{{\tt hep-th/9802150}}].

\bibitem{Gubser:1998bc}
S.~S. Gubser, I.~R. Klebanov and A.~M. Polyakov, \emph{{Gauge theory
  correlators from noncritical string theory}},
  \href{http://dx.doi.org/10.1016/S0370-2693(98)00377-3}{\emph{Phys. Lett.}
  {\bf B428} (1998) 105--114},
  [\href{https://arxiv.org/abs/hep-th/9802109}{{\tt hep-th/9802109}}].

\bibitem{Ryu:2006bv}
S.~Ryu and T.~Takayanagi, \emph{{Holographic derivation of entanglement entropy
  from AdS/CFT}},
  \href{http://dx.doi.org/10.1103/PhysRevLett.96.181602}{\emph{Phys. Rev.
  Lett.} {\bf 96} (2006) 181602},
  [\href{https://arxiv.org/abs/hep-th/0603001}{{\tt hep-th/0603001}}].

\bibitem{Ryu:2006ef}
S.~Ryu and T.~Takayanagi, \emph{{Aspects of Holographic Entanglement Entropy}},
  \href{http://dx.doi.org/10.1088/1126-6708/2006/08/045}{\emph{JHEP} {\bf 08}
  (2006) 045}, [\href{https://arxiv.org/abs/hep-th/0605073}{{\tt
  hep-th/0605073}}].

\bibitem{Song:2016pwx}
W.~Song, Q.~Wen and J.~Xu, \emph{{Generalized Gravitational Entropy for Warped
  Anti–de Sitter Space}},
  \href{http://dx.doi.org/10.1103/PhysRevLett.117.011602}{\emph{Phys. Rev.
  Lett.} {\bf 117} (2016) 011602},
  [\href{https://arxiv.org/abs/1601.02634}{{\tt 1601.02634}}].

\bibitem{Song:2016gtd}
W.~Song, Q.~Wen and J.~Xu, \emph{{Modifications to Holographic Entanglement
  Entropy in Warped CFT}},
  \href{http://dx.doi.org/10.1007/JHEP02(2017)067}{\emph{JHEP} {\bf 02} (2017)
  067}, [\href{https://arxiv.org/abs/1610.00727}{{\tt 1610.00727}}].

\bibitem{Jiang:2017ecm}
H.~Jiang, W.~Song and Q.~Wen, \emph{{Entanglement Entropy in Flat Holography}},
  \href{http://dx.doi.org/10.1007/JHEP07(2017)142}{\emph{JHEP} {\bf 07} (2017)
  142}, [\href{https://arxiv.org/abs/1706.07552}{{\tt 1706.07552}}].

\bibitem{Wen:2018mev}
Q.~Wen, \emph{{Towards the generalized gravitational entropy for spacetimes
  with non-Lorentz invariant duals}},
  \href{http://dx.doi.org/10.1007/JHEP01(2019)220}{\emph{JHEP} {\bf 01} (2019)
  220}, [\href{https://arxiv.org/abs/1810.11756}{{\tt 1810.11756}}].

\bibitem{Vidal}
Y.~Chen and G.~Vidal, \emph{{Entanglement contour}}, {\emph{Journal of
  Statistical Mechanics: Theory and Experiment} {\bf 2014} (2014) P10011},
  [\href{https://arxiv.org/abs/1406.1471}{{\tt 1406.1471}}].

\bibitem{Casini:2011kv}
H.~Casini, M.~Huerta and R.~C. Myers, \emph{{Towards a derivation of
  holographic entanglement entropy}},
  \href{http://dx.doi.org/10.1007/JHEP05(2011)036}{\emph{JHEP} {\bf 05} (2011)
  036}, [\href{https://arxiv.org/abs/1102.0440}{{\tt 1102.0440}}].

\bibitem{Jiang:2019qvd}
H.~Jiang, \emph{{Anomalous Gravitation and its Positivity from Entanglement}},
  \href{http://dx.doi.org/10.1007/JHEP10(2019)283}{\emph{JHEP} {\bf 10} (2019)
  283}, [\href{https://arxiv.org/abs/1906.04142}{{\tt 1906.04142}}].

\bibitem{Botero}
A.~Botero and B.~Reznik, \emph{Spatial structures and localization of vacuum
  entanglement in the linear harmonic chain},
  \href{http://dx.doi.org/10.1103/PhysRevA.70.052329}{\emph{Phys. Rev. A} {\bf
  70} (Nov, 2004) 052329}.

\bibitem{PhysRevB.92.115129}
I.~Fr\'erot and T.~Roscilde, \emph{Area law and its violation: A microscopic
  inspection into the structure of entanglement and fluctuations},
  \href{http://dx.doi.org/10.1103/PhysRevB.92.115129}{\emph{Phys. Rev. B} {\bf
  92} (Sep, 2015) 115129}.

\bibitem{Coser:2017dtb}
A.~Coser, C.~De~Nobili and E.~Tonni, \emph{{A contour for the entanglement
  entropies in harmonic lattices}},
  \href{http://dx.doi.org/10.1088/1751-8121/aa7902}{\emph{J. Phys.} {\bf A50}
  (2017) 314001}, [\href{https://arxiv.org/abs/1701.08427}{{\tt 1701.08427}}].

\bibitem{Tonni:2017jom}
E.~Tonni, J.~Rodriguez-Laguna and G.~Sierra, \emph{{Entanglement hamiltonian
  and entanglement contour in inhomogeneous 1D critical systems}},
  \href{http://dx.doi.org/10.1088/1742-5468/aab67d}{\emph{J. Stat. Mech.} {\bf
  1804} (2018) 043105}, [\href{https://arxiv.org/abs/1712.03557}{{\tt
  1712.03557}}].

\bibitem{DiGiulio:2019lpb}
G.~Di~Giulio, R.~Arias and E.~Tonni, \emph{{Entanglement hamiltonians in 1D
  free lattice models after a global quantum quench}},
  \href{http://dx.doi.org/10.1088/1742-5468/ab4e8f}{\emph{J. Stat. Mech.} {\bf
  1912} (2019) 123103}, [\href{https://arxiv.org/abs/1905.01144}{{\tt
  1905.01144}}].

\bibitem{Kudler-Flam:2019nhr}
J.~Kudler-Flam, H.~Shapourian and S.~Ryu, \emph{{The negativity contour: a
  quasi-local measure of entanglement for mixed states}},
  \href{http://dx.doi.org/10.21468/SciPostPhys.8.4.063}{\emph{SciPost Phys.}
  {\bf 8} (2020) 063}, [\href{https://arxiv.org/abs/1908.07540}{{\tt
  1908.07540}}].

\bibitem{Kudler-Flam:2019oru}
J.~Kudler-Flam, I.~MacCormack and S.~Ryu, \emph{{Holographic entanglement
  contour, bit threads, and the entanglement tsunami}},
  \href{http://dx.doi.org/10.1088/1751-8121/ab2dae}{\emph{J. Phys.} {\bf A52}
  (2019) 325401}, [\href{https://arxiv.org/abs/1902.04654}{{\tt 1902.04654}}].

\bibitem{MacCormack:2020auw}
I.~MacCormack, M.~T. Tan, J.~Kudler-Flam and S.~Ryu, \emph{{Operator and
  entanglement growth in non-thermalizing systems: many-body localization and
  the random singlet phase}},  \href{https://arxiv.org/abs/2001.08222}{{\tt
  2001.08222}}.

\bibitem{Swingle:2009bg}
B.~Swingle, \emph{{Entanglement Renormalization and Holography}},
  \href{http://dx.doi.org/10.1103/PhysRevD.86.065007}{\emph{Phys. Rev.} {\bf
  D86} (2012) 065007}, [\href{https://arxiv.org/abs/0905.1317}{{\tt
  0905.1317}}].

\bibitem{Freedman:2016zud}
M.~Freedman and M.~Headrick, \emph{{Bit threads and holographic entanglement}},
  \href{http://dx.doi.org/10.1007/s00220-016-2796-3}{\emph{Commun. Math. Phys.}
  {\bf 352} (2017) 407--438}, [\href{https://arxiv.org/abs/1604.00354}{{\tt
  1604.00354}}].

\bibitem{Han:2019scu}
M.~Han and Q.~Wen, \emph{{Entanglement entropies from entanglement contour:
  annuli and spherical shells}},  \href{https://arxiv.org/abs/1905.05522}{{\tt
  1905.05522}}.

\bibitem{araki1970}
H.~Araki and E.~H. Lieb, \emph{Entropy inequalities}, {\emph{Comm. Math. Phys.}
  {\bf 18} (1970) 160--170}.

\bibitem{Headrick:2013zda}
M.~Headrick, \emph{{General properties of holographic entanglement entropy}},
  \href{http://dx.doi.org/10.1007/JHEP03(2014)085}{\emph{JHEP} {\bf 03} (2014)
  085}, [\href{https://arxiv.org/abs/1312.6717}{{\tt 1312.6717}}].

\bibitem{Lieb:1973cp}
E.~H. Lieb and M.~B. Ruskai, \emph{{Proof of the strong subadditivity of
  quantum-mechanical entropy}},
  \href{http://dx.doi.org/10.1063/1.1666274}{\emph{J. Math. Phys.} {\bf 14}
  (1973) 1938--1941}.

\bibitem{Lieb:1973zz}
E.~H. Lieb and M.~B. Ruskai, \emph{{A Fundamental Property of
  Quantum-Mechanical Entropy}},
  \href{http://dx.doi.org/10.1103/PhysRevLett.30.434}{\emph{Phys. Rev. Lett.}
  {\bf 30} (1973) 434--436}.

\bibitem{Hayden:2011ag}
P.~Hayden, M.~Headrick and A.~Maloney, \emph{{Holographic Mutual Information is
  Monogamous}}, \href{http://dx.doi.org/10.1103/PhysRevD.87.046003}{\emph{Phys.
  Rev.} {\bf D87} (2013) 046003}, [\href{https://arxiv.org/abs/1107.2940}{{\tt
  1107.2940}}].

\bibitem{Hubeny:2007re}
V.~E. Hubeny and M.~Rangamani, \emph{{Holographic entanglement entropy for
  disconnected regions}},
  \href{http://dx.doi.org/10.1088/1126-6708/2008/03/006}{\emph{JHEP} {\bf 03}
  (2008) 006}, [\href{https://arxiv.org/abs/0711.4118}{{\tt 0711.4118}}].

\bibitem{Agon:2018lwq}
C.~A. Agón, J.~De~Boer and J.~F. Pedraza, \emph{{Geometric Aspects of
  Holographic Bit Threads}},
  \href{http://dx.doi.org/10.1007/JHEP05(2019)075}{\emph{JHEP} {\bf 05} (2019)
  075}, [\href{https://arxiv.org/abs/1811.08879}{{\tt 1811.08879}}].

\bibitem{Hubeny:2013gta}
V.~E. Hubeny, H.~Maxfield, M.~Rangamani and E.~Tonni, \emph{{Holographic
  entanglement plateaux}},
  \href{http://dx.doi.org/10.1007/JHEP08(2013)092}{\emph{JHEP} {\bf 08} (2013)
  092}, [\href{https://arxiv.org/abs/1306.4004}{{\tt 1306.4004}}].

\bibitem{Detournay:2012pc}
S.~Detournay, T.~Hartman and D.~M. Hofman, \emph{{Warped Conformal Field
  Theory}}, \href{http://dx.doi.org/10.1103/PhysRevD.86.124018}{\emph{Phys.
  Rev.} {\bf D86} (2012) 124018}, [\href{https://arxiv.org/abs/1210.0539}{{\tt
  1210.0539}}].

\bibitem{Compere:2013bya}
G.~Compère, W.~Song and A.~Strominger, \emph{{New Boundary Conditions for
  AdS3}}, \href{http://dx.doi.org/10.1007/JHEP05(2013)152}{\emph{JHEP} {\bf 05}
  (2013) 152}, [\href{https://arxiv.org/abs/1303.2662}{{\tt 1303.2662}}].

\bibitem{Balasubramanian:2013lsa}
V.~Balasubramanian, B.~D. Chowdhury, B.~Czech, J.~de~Boer and M.~P. Heller,
  \emph{{Bulk curves from boundary data in holography}},
  \href{http://dx.doi.org/10.1103/PhysRevD.89.086004}{\emph{Phys. Rev.} {\bf
  D89} (2014) 086004}, [\href{https://arxiv.org/abs/1310.4204}{{\tt
  1310.4204}}].

\bibitem{Czech:2014ppa}
B.~Czech and L.~Lamprou, \emph{{Holographic definition of points and
  distances}}, \href{http://dx.doi.org/10.1103/PhysRevD.90.106005}{\emph{Phys.
  Rev.} {\bf D90} (2014) 106005}, [\href{https://arxiv.org/abs/1409.4473}{{\tt
  1409.4473}}].

\bibitem{Czech:2015qta}
B.~Czech, L.~Lamprou, S.~McCandlish and J.~Sully, \emph{{Integral Geometry and
  Holography}}, \href{http://dx.doi.org/10.1007/JHEP10(2015)175}{\emph{JHEP}
  {\bf 10} (2015) 175}, [\href{https://arxiv.org/abs/1505.05515}{{\tt
  1505.05515}}].

\bibitem{Czech:2015kbp}
B.~Czech, L.~Lamprou, S.~McCandlish and J.~Sully, \emph{{Tensor Networks from
  Kinematic Space}},
  \href{http://dx.doi.org/10.1007/JHEP07(2016)100}{\emph{JHEP} {\bf 07} (2016)
  100}, [\href{https://arxiv.org/abs/1512.01548}{{\tt 1512.01548}}].

\bibitem{Abt:2018ywl}
R.~Abt, J.~Erdmenger, M.~Gerbershagen, C.~M. Melby-Thompson and C.~Northe,
  \emph{{Holographic Subregion Complexity from Kinematic Space}},
  \href{http://dx.doi.org/10.1007/JHEP01(2019)012}{\emph{JHEP} {\bf 01} (2019)
  012}, [\href{https://arxiv.org/abs/1805.10298}{{\tt 1805.10298}}].

\bibitem{Castro:2015csg}
A.~Castro, D.~M. Hofman and N.~Iqbal, \emph{{Entanglement Entropy in Warped
  Conformal Field Theories}},
  \href{http://dx.doi.org/10.1007/JHEP02(2016)033}{\emph{JHEP} {\bf 02} (2016)
  033}, [\href{https://arxiv.org/abs/1511.00707}{{\tt 1511.00707}}].

\bibitem{Barnich:2010eb}
G.~Barnich and C.~Troessaert, \emph{{Aspects of the BMS/CFT correspondence}},
  \href{http://dx.doi.org/10.1007/JHEP05(2010)062}{\emph{JHEP} {\bf 05} (2010)
  062}, [\href{https://arxiv.org/abs/1001.1541}{{\tt 1001.1541}}].

\bibitem{Bagchi:2010eg}
A.~Bagchi, \emph{{Correspondence between Asymptotically Flat Spacetimes and
  Nonrelativistic Conformal Field Theories}},
  \href{http://dx.doi.org/10.1103/PhysRevLett.105.171601}{\emph{Phys. Rev.
  Lett.} {\bf 105} (2010) 171601}, [\href{https://arxiv.org/abs/1006.3354}{{\tt
  1006.3354}}].

\bibitem{Bagchi:2012cy}
A.~Bagchi and R.~Fareghbal, \emph{{BMS/GCA Redux: Towards Flatspace Holography
  from Non-Relativistic Symmetries}},
  \href{http://dx.doi.org/10.1007/JHEP10(2012)092}{\emph{JHEP} {\bf 10} (2012)
  092}, [\href{https://arxiv.org/abs/1203.5795}{{\tt 1203.5795}}].

\bibitem{Bagchi:2014iea}
A.~Bagchi, R.~Basu, D.~Grumiller and M.~Riegler, \emph{{Entanglement entropy in
  Galilean conformal field theories and flat holography}},
  \href{http://dx.doi.org/10.1103/PhysRevLett.114.111602}{\emph{Phys. Rev.
  Lett.} {\bf 114} (2015) 111602}, [\href{https://arxiv.org/abs/1410.4089}{{\tt
  1410.4089}}].

\bibitem{Basu:2015evh}
R.~Basu and M.~Riegler, \emph{{Wilson Lines and Holographic Entanglement
  Entropy in Galilean Conformal Field Theories}},
  \href{http://dx.doi.org/10.1103/PhysRevD.93.045003}{\emph{Phys. Rev.} {\bf
  D93} (2016) 045003}, [\href{https://arxiv.org/abs/1511.08662}{{\tt
  1511.08662}}].

\bibitem{Hijano:2017eii}
E.~Hijano and C.~Rabideau, \emph{{Holographic entanglement and Poincaré blocks
  in three-dimensional flat space}},
  \href{http://dx.doi.org/10.1007/JHEP05(2018)068}{\emph{JHEP} {\bf 05} (2018)
  068}, [\href{https://arxiv.org/abs/1712.07131}{{\tt 1712.07131}}].

\bibitem{Takayanagi:2017knl}
T.~Takayanagi and K.~Umemoto, \emph{{Entanglement of purification through
  holographic duality}},
  \href{http://dx.doi.org/10.1038/s41567-018-0075-2}{\emph{Nature Phys.} {\bf
  14} (2018) 573--577}, [\href{https://arxiv.org/abs/1708.09393}{{\tt
  1708.09393}}].

\bibitem{Nguyen:2017yqw}
P.~Nguyen, T.~Devakul, M.~G. Halbasch, M.~P. Zaletel and B.~Swingle,
  \emph{{Entanglement of purification: from spin chains to holography}},
  \href{http://dx.doi.org/10.1007/JHEP01(2018)098}{\emph{JHEP} {\bf 01} (2018)
  098}, [\href{https://arxiv.org/abs/1709.07424}{{\tt 1709.07424}}].

\bibitem{Kudler-Flam:2018qjo}
J.~Kudler-Flam and S.~Ryu, \emph{{Entanglement negativity and minimal
  entanglement wedge cross sections in holographic theories}},
  \href{http://dx.doi.org/10.1103/PhysRevD.99.106014}{\emph{Phys. Rev.} {\bf
  D99} (2019) 106014}, [\href{https://arxiv.org/abs/1808.00446}{{\tt
  1808.00446}}].

\bibitem{Faulkner:2017vdd}
T.~Faulkner and A.~Lewkowycz, \emph{{Bulk locality from modular flow}},
  \href{http://dx.doi.org/10.1007/JHEP07(2017)151}{\emph{JHEP} {\bf 07} (2017)
  151}, [\href{https://arxiv.org/abs/1704.05464}{{\tt 1704.05464}}].

\bibitem{Wen:2019iyq}
Q.~Wen, \emph{{Formulas for Partial Entanglement Entropy}},
  \href{https://arxiv.org/abs/1910.10978}{{\tt 1910.10978}}.

\end{thebibliography}\endgroup

\end{document}